\documentclass[12pt]{article}
\pdfoutput=1

\usepackage{cite}
\usepackage{booktabs}
\usepackage[english]{babel}
\usepackage{amsmath,amssymb,amsbsy,amstext, amsthm, simplewick}
\usepackage{hyperref}
\usepackage{graphicx}
\usepackage{amsfonts}
\usepackage{amssymb}
\usepackage[small]{caption}
\usepackage{upgreek}
\usepackage[svgnames,dvipsnames,x11names,table]{xcolor}
\usepackage{multirow}
\usepackage{geometry}
\usepackage[hang,flushmargin]{footmisc}
\usepackage{bm}
\usepackage{braket}
\usepackage{subcaption}
\usepackage{mathtools}
\usepackage{setspace}
\usepackage{cleveref}
\usepackage{comment}
\usepackage{scalerel}
\usepackage[normalem]{ulem}
\usepackage{slashed}
\usepackage{enumitem}
\usepackage{dsfont}
\usepackage{tikz}
\usetikzlibrary{decorations.markings}
\usetikzlibrary{shapes.misc}

\makeatletter
\g@addto@macro\bfseries{\boldmath}
\makeatother

\hypersetup{
    colorlinks=true,
    linkcolor={red!50!black},
    citecolor={blue!50!black},
    urlcolor={blue!80!black}
}

\usepackage{colortbl}

\makeatletter
\newlength{\apb@width}
\newcommand{\autoparbox}[2][c]{\settowidth{\apb@width}{#2}\parbox[#1]{\apb@width}{#2}}

\makeatother

\definecolor{lightgray}{gray}{0.9}

\usepackage[framemethod=default]{mdframed}
\newmdenv[skipabove=7pt,
skipbelow=7pt,
rightline=false,
leftline=false,
topline=false,
bottomline=false,
backgroundcolor=gray!10,
linecolor=gray,
innerleftmargin=5pt,
innerrightmargin=5pt,
innertopmargin=5pt,
innerbottommargin=5pt,
leftmargin=0cm,
rightmargin=0cm,
linewidth=4pt]{eBox}

\usepackage[most]{tcolorbox}
\tcbset{colback=white, colframe=black,
        highlight math style= {enhanced, 
            colframe=red,colback=red!10!white,boxsep=0pt}
        }
\definecolor{light-gray}{gray}{0.95}

\crefname{table}{Table}{Tables}
\crefname{equation}{Eq.}{Eqs.}
\crefname{appendix}{App.}{Apps.}
\crefname{section}{Sec.}{Secs.}
\crefname{figure}{Fig.}{Figs.}

\numberwithin{equation}{section}

\def\beq{\begin{equation}}
\def\eeq{\end{equation}}

\def\bea{\begin{eqnarray}}
\def\eea{\end{eqnarray}}

\def\beq{\begin{equation}}
\def\eeq{\end{equation}}
\def\bea{\begin{eqnarray}}
\def\eea{\end{eqnarray}}

\def\O{{\cal O}}

\def\Mpl{M_{\rm pl}}
\def\Mp{M_{\rm pl}}

\def\k{{\vec{\scaleto{k}{7pt}}}}

\def\x{{\vec x}}

\DeclareRobustCommand{\SkipTocEntry}[4]{}

\definecolor{colorTC}{rgb}{.2,.7,.2}
\definecolor{amethyst}{rgb}{0.6, 0.4, 0.8}

\definecolor{acolor}{rgb}{0.4, 0.2, 0.4}

\definecolor{blue3}{RGB}{31, 119, 180}
\definecolor{red3}{RGB}{	214, 39, 40}
\definecolor{orange3}{RGB}{255, 127, 14}
\definecolor{green3}{RGB}{44, 160, 44}

\begin{document}

\begin{titlepage}
\setcounter{page}{1} \baselineskip=15.5pt
\thispagestyle{empty}
$\quad$
\vskip 70 pt

\begin{center}
{\fontsize{18}{18} \bf A Goldstone Boson Equivalence for Inflation}
\end{center}

\vskip 20pt
\begin{center}
\noindent
{\fontsize{12}{18}\selectfont  Daniel Green, Kshitij Gupta, and Yiwen Huang}
\end{center}

\begin{center}
\vskip 4pt
\textit{{\small Department of Physics, University of California at San Diego,  La Jolla, CA 92093, USA}}

\end{center}

\vspace{0.4cm}
 \begin{center}{\bf Abstract}
 \end{center}

\noindent
The effective field theory of single-field inflation characterizes the inflationary epoch in terms of a pattern of symmetry breaking. An operator acquires a time-dependent vacuum expectation value, defining a preferred spatial slicing. In the absence of dynamical gravity, the fluctuations around the time-dependent background are described by the Goldstone boson associated with this symmetry breaking process. With gravity, the Goldstone is eaten by the metric, becoming the scalar metric fluctuation. In this paper, we will show that in general single-field inflation, the statistics of scalar metric fluctuations are given by the statistics of this Goldstone boson {\it decoupled from gravity} up to corrections that are controlled as an expansion in slow-roll parameters. This even holds in the presence of additional parameters, like the speed of sound, that naively enhance the impact of the gravitational terms. In the process, we derive expressions for leading and sub-leading gravitational corrections to all-orders in the Goldstone boson.

\end{titlepage}

\setcounter{page}{2}

\restoregeometry

\begin{spacing}{1.2}
\newpage
\setcounter{tocdepth}{2}
\tableofcontents
\end{spacing}

\setstretch{1.1}
\newpage

\section{Introduction}

One of the central questions in inflationary cosmology is whether single- or multi-field inflation is needed to explain observed structure in the universe~\cite{Achucarro:2022qrl}. Current observations are consistent with both~\cite{Planck:2019kim}, but future surveys offer the potential to distinguish these two classes of ideas~\cite{SPHEREx:2014bgr,DESI:2022lza,Schlegel:2022vrv,Chang:2022lrw}. The single-field consistency conditions~\cite{Maldacena:2002vr,Creminelli:2004yq} are essential in this quest to understand the nature of inflation. These arise when the inflaton is the only relevant degree of freedom for the formation of structure; all scalar fluctuations can be absorbed (are ``eaten") by the metric, meaning they are subject to the constraints of diffeomorphism invariance like any other observable of the metric~\cite{Bardeen:1980kt,Salopek:1990jq}. With multiple degrees of freedom, there are non-metric scalar fluctuations in every gauge~\cite{Lyth:2002my,Zaldarriaga:2003my,Sasaki:2006kq,Chen:2009zp,Senatore:2010wk,Baumann:2011nk}, giving rise to new kinds of signatures often broadly described as ``cosmological collider physics"~\cite{Arkani-Hamed:2015bza}.

While it is of critical importance that adiabatic fluctuations can be entirely absorbed into the metric, practical calculations of cosmological correlators are often more cumbersome in this description. In many circumstances, the largest contributions to the scalar metric fluctuations can be calculated accurately in terms of the fluctuations of the inflaton~\cite{Arkani-Hamed:2003juy,Alishahiha:2004eh,Chen:2006nt}, and other fields, on a non-dynamical background. However, the evolution of the universe couples the metric and field theoretic degrees of freedom. Ultimately, any complete calculation of inflationary statistics involves a nontrivial manipulation of the action to incorporate the constraints imposed by gravity~\cite{Maldacena:2002vr}.

Much of these properties were formalized in the context of the EFT of inflation~\cite{Creminelli:2006xe,Cheung:2007st}. Models of single-field inflation are characterized by having spontaneously broken time translations, resulting in a Goldstone boson, $\pi$. Gravity gauges the time translation and eats the Goldstone boson, as described above. However, like the case of a massive gauge boson~\cite{Cornwall:1974km,Lee:1977eg,Chanowitz:1985hj,Gounaris:1986cr,Yao:1988aj,Bagger:1989fc,Veltman:1989ud}, we should recover the Goldstone boson description in the limit where dynamical gravity is decoupled (i.e.~a Goldstone boson equivalence theorem). This was demonstrated directly in~\cite{Baumann:2011su}: given a Hubble parameter $H(t)$ and reduced Planck-mass $\Mpl$, the Goldstone boson decouples from gravity when we take $\Mpl^2 \to \infty$ and $\dot H \to 0$ while holding $H$ and $\Mpl^2 \dot H$ fixed. The Goldstone boson may still have non-trivial self-interactions in this limit, which are ultimately connected to large non-Gaussian correlators of the scalar metric fluctuations.

The existence of the decoupling limit suggests that the Goldstone boson description should be a good approximation to the scalar metric degree of freedom, up to corrections that would vanish in that limit. The general expectation is that these corrections are always suppressed by slow-roll parameters, e.g.~by $\epsilon \equiv - \dot H/H^2 \ll 1$. This would ensure the description in terms of a Goldstone boson on a fixed inflating background is sufficient for calculating non-Gaussian correlators in a wide range of models. This is what we will refer to as the Goldstone boson equivalence theorem for inflation. This intuition has been confirmed in a number of cases by direct calculation~\cite{Arkani-Hamed:2003juy,Alishahiha:2004eh,Chen:2006nt,Cheung:2007sv}.

Unfortunately, the decoupling limit alone does not constitute a sufficient proof of the equivalence theorem for many practical applications. Specifically, models of inflation contain other large dimensionless parameters, such as the inverse speed of sound, $c_s^{-1}$, which often multiply the slow-roll parameter. For example, in the original paper on the EFT of inflation~\cite{Cheung:2007st}, the mixing with gravity was estimated to scale as $\epsilon/c_s^2$. This vanishes in the decoupling limit $\epsilon\to 0$, but could be large in models where $c_s^2 < \epsilon$. Yet, direct calculations show that this is likely an over-estimate of the role of mixing; in the simplest small-$c_s$ models, the gravitational corrections to the cosmological correlators at first order in the amplitude of the metric fluctuations are proportional to $\epsilon$ with no additional $c_s$-factors~\cite{Chen:2006nt,Cheung:2007sv}. 

The small size of some of these corrections can be understood from the long-wavelength description in terms of the metric fluctuation, $\zeta$, which is necessarily time-independent outside the horizon~\cite{Salopek:1990jq}. This behavior has been proven to all orders in perturbation theory~\cite{Senatore:2012ya,Assassi:2012et,Cohen:2020php}, and fixes the form of any non-derivative gravitational corrections to the Goldstone boson action, or alternatively the soft behavior of correlation function~\cite{Weinberg:2003sw,Creminelli:2012ed,Hinterbichler:2012nm,Hinterbichler:2013dpa,Pajer:2013ana}. However, this is only a subset of the possible gravitational corrections and thus is insufficient for a general proof. Alternatively, it has been shown the high-energy scattering of the inflaton combined with the soft behavior of $\zeta$ is sufficient to fix the form of the three-point function~\cite{Pajer:2020wxk}. This is also insufficient to prove a Goldstone boson equivalence, as one would again need to show the scattering amplitudes can be calculated without gravity~\cite{Arkani-Hamed:2018kmz,Pajer:2020wxk,Jazayeri:2021fvk,Benincasa:2022gtd}.

The goal of this paper is to demonstrate that inflationary correlators can be calculated to leading order in slow-roll using the Goldstone boson action on a fixed inflationary geometry (i.e.~ignoring the mixing with gravity). Specifically, we will show that the mixing with gravity is always suppressed by slow-roll parameters and is not enhanced by model dependent factors like the speed of sound. In the process, we will provide general results for the first- and second-order slow-roll corrections to the action, to all orders in the Goldstone fluctuations. 

Our results make use of three observations that simplify the analysis. First, the leading coupling of gravity to Goldstone boson is determined by the stress tensor on a fixed Friedmann–Robertson–Walker (FRW) background. Second, the natural size of higher-dimension operators in the EFT are also enhanced by powers of $c_s$~\cite{Senatore:2010jy,Baumann:2015nta,Baumann:2019ghk,Green:2022slj}. We will find that gravitational corrections are only suppressed relative to operators in the decoupled action when the higher-dimensional couplings follow the expected $c_s$ scaling from naive-dimensional analysis. Finally, we will observe that the scaling with $c_s$ can be understood on dimensional grounds and thus is predictable at higher orders.

This paper is organized as follows: In Section~\ref{sec:review}, we review the EFT of inflation and the decoupling limit. In Section~\ref{sec:cs}, we demonstrate that dynamical gravitational corrections are slow-roll suppressed in the EFT of inflation at small sound speed. In Section~\ref{sec:generalizations}, we discuss the generalization to other non-relativistic dispersion relations for the Goldstone. We conclude in Section~\ref{sec:conclusions}. 

The paper has three appendices: In Appendix~\ref{app:ADM}, we review the explicit calculation of the ADM constraint to first order in slow roll and linear in the Goldstone boson. In Appendix~\ref{app:stress}, we derive some important results regarding the stress tensor in the EFT of inflation that are used throughout the paper. In Appendix~\ref{app:esp2}, we derive the second order corrections to the metric.

\section{The EFT of Inflation and Gravity}\label{sec:review}

In order to define our Goldstone boson equivalence, we will first need to review the EFT of inflation. We will present the EFT in the form most similar from the breaking of global symmetries and the Higgs mechanism. In this language, the meaning of the equivalence theorem in the inflationary context will be most straightforward and most similar to the equivalence theorem for massive gauge fields~\cite{Cornwall:1974km,Lee:1977eg,Chanowitz:1985hj,Gounaris:1986cr,Yao:1988aj,Bagger:1989fc,Veltman:1989ud}.

\subsection{Action}

The EFT of inflation is defined by a pattern of symmetry breaking. We presume there exists an operator $\O$ that acquires a time-dependent background $\langle \O \rangle \propto t$ such that time-translations of the (rigid) flat space limit would be spontaneously broken. Following the symmetry breaking playbook, we define a variable 
\beq
U \equiv t + \pi(t,\x) \ ,
\eeq
such that $U(\x,t)$ transforms linearly under time translation, $t\to t+c$, and therefore Goldstone boson $\pi$ transforms non-linearly, $\pi \to \pi -c$. We are (weakly) coupling this EFT to gravity, and the time-translations are gauged (making them a local time diffeomorphism). As a result, under a diffeomorphism $x^\mu \to x^\mu +\xi^\mu(\x,t)$, $\pi \to \pi -\xi^0$ so that $U(\x,t)$ transforms as a scalar.

The value of defining $U$ is that we can the write the most general action for $\pi$ on a curved background as
\beq
S_\pi = \int d^4 x \sqrt{-g} {\cal G}[U, \nabla_\mu]  \ ,
\eeq
where ${\cal G}$ any scalar function of $U(\x,t)$ and covariant derivatives thereof. Expanding in powers of $\pi$ and derivatives per field, we often write
\begin{align}
S_\pi &=\int d^4 x \sqrt{-g} \sum_{n=0}^{\infty}  \frac{M^4_n(U)}{n!}  (\partial_\mu U \partial^\mu U+1)^n + \ldots \\
&=\int d^4 x \sqrt{-g} \sum_{n=0}^{\infty} \frac{M^4_n(t+\pi)}{n!}  (-2\dot \pi + \partial_\mu \pi \partial^\mu \pi)^n + \ldots \ ,\label{eq:Mn_action}
\end{align}
where $\ldots$ contains terms with more than one derivative per field or higher and $\dot \pi = \frac{\partial}{\partial t} \pi$. The additional $+1$ with each term is a convention chosen so that $M_n^4$ comes at order ${\cal O}(\pi^n)$.

This action is subject to two constraints: (1) $\pi = 0$ should be a solution to the equations of motion (no tadpoles) and (2) when coupled to gravity, we should have a homogeneous FRW solution for an $a(t)$ consistent with an inflationary background. These requirements are related and are determined by the equations of motion of the action, 
\begin{align}
S &= S_{\rm EH} + S_\pi  \\
&=\int d^4 x \sqrt{-g} \left[ \frac{\Mp^2}{2} R +M_0^4(U) + M_1^4(U)(\partial_\mu U \partial^\mu U+1) + {\cal O}(\pi^2) \right]\ .
\end{align}
Assuming $g_{\mu \nu}$ has a flat FRW solution when $\pi= 0$ gives
\bea
3 \Mp^2 H^2 &=& -M_0^4(t)-2 M_1^4(t)  \\
3 \Mp^2 (\dot H + H^2) &=& -M_0^4(t)+ M_1^4(t) 
\eea
where $H(t) = \dot a(t) /a(t)$. Combing these equations yields
\beq
M_1^4= \Mp^2 \dot H \qquad M_0^4 = - \Mp^2(3H^2 + 2 \dot H) \ . 
\eeq
As a result, the action of the effective theory of inflation is given by 
\bea\label{eq:EFTaction}
S = \int d^4 x \sqrt{-g} [ \, \tfrac{1}{2} M_{\rm pl}^2 R- \Mp^2(3 H^2 + \Mp^2 \dot H)+\Mp^2 \dot H \partial_\mu (t+\pi) \partial^\mu(t+\pi) +\ldots]\ .
\eea
One can check that linear term in $\pi$ is a total derivative. We did not need to impose the absence of a tadpole separately as it is redundant with Einstein's equations through the conservation of the stress tensor.

\subsection{The Decoupling Limit}

One of the key ideas in the construction of the EFT of inflation is that, in the subhorizon limit, the dynamics of the inflaton look like the spontaneous breaking of time-translations in flat space. As such, the scalar density fluctuation should encode the would-be Goldstone boson. However, like any gauged symmetry, this is just a story, as the metric eats the Goldstone. However, to be consistent, we expect that if we decouple dynamical gravity, the Goldstone description emerges. This idea is also central to the Goldstone boson equivalence theorem~\cite{Cornwall:1974km,Lee:1977eg,Chanowitz:1985hj,Gounaris:1986cr,Yao:1988aj,Bagger:1989fc,Veltman:1989ud}.

The naive decoupling limit is to simply take the gravitational constant $G_N \to 0$ or (equivalently) $\Mpl \to \infty$. However, we have also imposed that our background is (a) a solution to Einstein's equation and (b) is curved. As a result, taking $\Mpl \to \infty$ holding $H$, $\dot H$ fixed is not necessarily holding fixed the physics of the matter sector of interest.

The mistake we are making is that we want to decouple dynamical gravity while maintaining both the action of the Goldstone boson and a de Sitter (rather than flat) background metric. The correct choice to accomplish this goal is 
\beq
\Mpl^2 \to \infty \qquad \dot H \to 0 \qquad H, \Mpl^2 \dot H, M_{n>1}^4= {\rm constant}  \ .
\eeq
The choice to hold $M_n^4$ constant is a simplifying assumption that the Goldstone boson action is time independent in the decoupling limit. 

The need for the condition $\Mpl^2 \to \infty$ should be clear, as it decouples the dynamical metric fluctuations. Taking $\dot H \to 0$ simultaneously, while holding $\Mpl^2 \dot H$ constant, is required to maintain the normalization of the relativistic kinetic term for $\pi$, 
\beq
S_2 =\int d^4 x \sqrt{-g} \Mp^2 \dot H \partial_\mu \pi \partial^\mu\pi \ ,
\eeq
and therefore we are holding this fixed. However, this may seem like an unphyiscal condition as it is just the normalization of the kinetic term. Instead, we can appeal directly to the idea that $\pi$ is the Goldstone boson of spontanous symmetry breaking and $T_{00}$ is the charge density associated with that symmetry. When $M_{2}^4= 0$, we have
\beq
T_{00} = - 2 \Mpl^2 \dot H \dot \pi + {\cal O}(\pi^2)
\eeq
such that
\beq
f_\pi^4 = -2 \Mpl^2 \dot H \ ,
\eeq
is the decay constant that defines the scale of symmetry breaking. In single-field inflation, $-2\Mpl^2 \dot H = \dot \phi^2$ as one would expect for the scale of spontaneously broken time translations. Our decoupling limit is therefore holding this scale fixed and decoupling dynamical gravity.

Holding $H$ fixed does require that the term $3 \Mpl H^2 \to \infty$ in Equation~(\ref{eq:EFTaction}). Since $\dot H \to 0$, this is just the cosmological constant that is necessary to maintain the background when we take $\Mpl \to \infty$, but does not introduce any large numbers into the dynamics of the scalar fluctuations.

Holding the remaining $M_{n>1}^4(t)$ parameters fixed ensures that we are not changing the action for $\pi$ in the limit where the metric is fixed. Physically, this means the any surviving self-interactions of $\pi$ are the consequence of scalar dynamics directly, rather than being induced by the gravitational sector. However, this distinction is not entirely universal, as the matching to the UV description that determined $M_n(t)$ may include contributions from integrating-out fields coupled to the inflaton by gravity.

The most important non-universal parameter of the EFT of inflation is $M^4_2$. It controls the speed of propagation of the scalar fluctuations (the speed of sound), $c_s$. Defining
\beq
M_2 = -2\frac{\Mpl^2 \dot H}{c_s^2} (1-c_s^2)
\eeq
one finds the quadratic action for the Goldstone boson becomes 
\beq
S_2 =\int d^4 x \sqrt{-g} \frac{\Mp^2 \dot H}{c_s^2} \left( \dot \pi^2 - \frac{c_s^2}{a^2}  \partial_i \pi \partial^i \pi \right) \ .
\eeq
In the presence of $c_s <1$, the stress tensor is similarly changed
\beq
T_{00} = - 2  \frac{\Mpl^2 \dot H}{c_s^2} \dot \pi + {\cal O}(\pi^2) \ .
\eeq
To determine the decay constant, we note that $T_{00}$ is an energy density but we want $f_\pi$ to be an energy. Using $c_s$ to account for this difference in units, one finds the $c_s$ generalization 
\beq
f_\pi^4 = -2 \Mpl^2 \dot H c_s \ .
\eeq
The powers of $c_s$ in this expression highlight our challenge in making sense of gravitational corrections. In this case, the inverse powers of $c_s$ in kinetic term are a simple consequence of units. On the other hand, when $c_s \ll 1$, it is useful to define  
\beq
M_n^4 \equiv c_n \frac{f_\pi^4}{c_s^{2n-1}}
\eeq
where $c_n = {\cal O}(1)$. Here $c_s$ appears not because of units, but because the strong coupling scale (which controls the interactions) is $\Lambda \approx f_\pi c_s$. The scaling of $M_n^4$ ensures that all these operators are controlled by this one scale.

The fact that we held the curved metric fixed in the decoupling limit, rather than taking the flat space limit simultaneously, further implies that the amplitude of scalar fluctuations is also held fixed. Concretely, since $\dot H \to 0$, the geometry is pure dS and the adiabatic fluctuation $\zeta \approx - H \pi$ has the power spectrum
\beq
\langle \zeta(\k ) \zeta(\k')\rangle = \frac{H^4}{2 f_\pi^4} \frac{1}{k^3} (2\pi)\delta(\k+\k')
\eeq
From the measured amplitude $A_s = H^4/(4\pi^2 f_\pi^4) = 2.2 \times 10^{-9}$~\cite{Planck:2018vyg}, we can infer~\cite{Planck:2018jri} that $f_\pi = 59 H$. This expression is valid for any $c_s$ and demonstrates the utility of thinking in terms of the scale $f_\pi$ for the physics of the Goldstone.

Intuitively, the decoupling limit suggests the following strong form of the Goldstone boson equivalence theorem: the scalar fluctuations in single-field inflation are determined by a decouple Goldstone boson $\pi$, up to corrections that are suppressed by 
\beq
\frac{\dot H}{H^2} \ll 1 \qquad \frac{H^2}{\Mp^2} \ll 1 \ .
\eeq
The key observation here is that the correction must vanish in the decoupling limit and thus must be proportional to one of these small parameters. However, being consistent with the decoupling limit does not determine where powers of $c_s$ might appear or whether these terms are large when $\Mpl$ is finite\footnote{We will not explore situations, such as primordial features~\cite{Slosar:2019gvt}, where the higher-order slow-roll parameters are large. These situations have unique theoretical and phenomenological behavior~\cite{Chen:2006xjb,Chen:2008wn,Flauger:2009ab,Flauger:2010ja,Behbahani:2011it,Behbahani:2012be,Flauger:2014ana,Flauger:2016idt} that requires a separate treatment that is beyond the scope of this work.}. This subtlety, along with the fact that we are working in a cosmological background with $H > 0$, makes the inflationary Goldstone boson equivalence distinct from its vector boson analogue. Nevertheless, some aspects of the EFT of inflation are better articulated in the flat space limit, $H \to 0$, where quantities like the S-matrix become useful~\cite{Baumann:2015nta,Pajer:2018egx,Pajer:2020wnj,Green:2022fwg,Hui:2022dnm,Green:2022slj,Cheung:2023qwn}.

\subsection{Choice of Gauge}

In reality, working at finite $\Mpl$ means that the time-translations are gauged. Rather than a spontaneosly broken time-translation symmetry, the theory has a redundancy under diffeomorphisms of the time coordinate. Recall that under a general diffeomorphism, the coordinates transform as $x^\mu \to x^\mu+ \xi^\mu(\x,t)$. In order for $U$ to transform as a scalar under a diffeomorphism, $\pi$ must transform non-linearly, 
\beq
\pi \to \pi - \xi^0(x) \ .
\eeq
As a result, the meaning of $\pi$ is ambiguous once we introduce gravity.

Before we can discuss the Goldstone boson equivalence theorem, we must address the relationship between gauge dependent quantities and some ``gauge-invariant" observable~\cite{Bardeen:1980kt}. For our purposes, there are two important choices of gauge that highlight the dynamics of the EFT of inflation:
\vskip 4pt
{\bf $\pi$-gauge :} In this gauge, the propagating degree of freedom in the action is entirely in the $\pi$, up to constraints from gravity. Specifically, we take the metric to be of the form
\beq
ds^2 = - N^2 dt^2 + a(t)^2\delta_{ij} (d x^i + N^i dt) (d x^j + N^j dt) \ .
\eeq
and places no constraint on $\pi$.

\vskip 4pt
{\bf $\zeta$-gauge :} 
also called {\bf unitary gauge}, this gauge is defined by setting $\pi(\x,t) =0$ such that all fluctuations appear in the metric
\beq
ds^2 = - N^2 dt^2 + a(t)^2 e^{2 \zeta(\x,t)} \delta_{ij} (d x^i + N^i dt) (d x^j + N^j dt) \ .
\eeq
In the context of an inflationary model driven by a scalar field $\phi(x,t)$, in this gauge our time variable $t$ defines surfaces of constant $\phi$ and $\zeta(\x)$ in the metric is the same as the Bardeen variable (which is gauge invariant)~\cite{Bardeen:1980kt,Salopek:1990jq}.

When changing gauge, one must remember that if we start in a set of coordinates $x^\mu$ with $\pi(x)\neq 0$ and $\zeta=0$ and we change to coordinates $\tilde x^\mu$ where $\tilde \zeta(\tilde x)\neq 0$ and $\tilde \pi(\tilde x) = 0$.  As a result, we transformation $\pi(x) = \tilde \pi (\tilde x) + \xi^0 (\tilde x)$ so that $\tilde \pi(\tilde x) = 0$. To find the correct $\tilde x = x + \xi$, we solve
\beq
\pi(\tilde x - \xi(\tilde x) ) = \xi^0(\tilde x) \ .
\eeq
This is easier to understand as follows, we want
\beq
\tilde t + \tilde \pi(\tilde t) = t + \pi(t)
\eeq
so if $\tilde t = t - \xi^0$ then this equation means that
\beq
\pi( t = \tilde t- \xi^0) =\tilde \pi(\tilde t) +\xi^0 (\tilde t)\ .
\eeq
Demanding $\tilde \pi(\tilde t) = 0$ is what determines $\xi$.

Since $\pi$ is small, we may solve order by order.  For example, to third order, we have
\beq
\xi^0 = \pi - \dot \pi \pi + \tfrac{1}{2} \ddot \pi \pi^2+ \dot \pi^2 \pi \ .
\eeq
The complication in matching to the metric is the following.  We want to write the metric in $\zeta$ gauge $\tilde g_{\mu \nu}$ via the diffeomorphism $ g_{\mu \nu} (x(\tilde x)) = \partial_{\mu} \tilde x^{\lambda}  \partial_{\nu} \tilde x^{\rho} \tilde g_{\lambda \rho}(\tilde x)$ .  In this case, using
\beq
ds^2 = - N^2 dt^2 + a(t)^2\delta_{ij} (d x^i + N^i dt) (d x^j + N^j dt) \ .
\eeq
we find after a diffeomorphism that 
\beq
\tilde g_{ij} = a^2(t-\xi^0) (\delta_{ij} - N_i \partial_j \xi^0 - N_j \partial_i \xi^0 ) +(a^2(t-\xi^0) N^i N_i- N^2 )\partial_i \xi^0 \partial_j \xi^0
\eeq
This is not in the Arnowitt-Deser-Misner (ADM) form, even in $\zeta$ gauge.  Therefore, we want to do a spatial diff to return to that gauge.  This is quite challenging, bur fortunately, is also unnecessary. Instead we notice that all the problems are proportional to $\partial_i T$.  In the limit $k \to 0$, these are subdominant contributions to $\zeta$, so as long as we match well after horizon crossing, it is easy to compute the effect.  We define the metric in $\zeta$ gauge as 
\beq
ds^2 = - N^2 dt^2 + a(t)^2\delta_{ij} e^{2 \zeta} (d x^i + N^i dt) (d x^j N^j dt) \ ,
\eeq
such that in the $ k \to 0$ limit, 
\beq
\zeta = \log \frac{ a(t - \xi^{0}) }{a(t)} = - H \pi +\tfrac{1}{2} \dot H \pi^2 + H \dot \pi \pi+ \ldots
\eeq
In a certain trivial sense, this is already a Goldstone boson equivalence statement. The scalar metric mode, $\zeta$, can be written entirely in terms of the dynamics of a Goldstone boson, $\pi$. However, it is very important that $\pi$ in this expression is still coupled to gravity. The complete answer for any correlator of $\zeta$ calculate in the $\pi$-gauge will require the same careful treatment of the gravitational constraints as it does in the $\zeta$ gauge. Nevertheless, the gauge transformation amounts to a weak statement of equivalence principle for inflation as identifies the precise relationship between the scalar metric fluctuations and the Goldstone boson, $\pi$. 

\subsection{ADM Constraints}

Choosing to work in $\pi$-gauge, we want to understand the origin of gravitational corrections, following~\cite{Chen:2006nt,Cheung:2007sv}. Although the dynamical degrees of freedom are expressed in terms of $\pi$, the metric is not entirely fixed. Instead, after fixing $\zeta =0$, we are left with the ADM decomposition
\beq
ds^2 = - N^2 dt^2 + a(t)^2\delta_{ij} (d x^i + N^i dt) (d x^j + N^j dt) \ .
\eeq
Taking derivatives of the action with respect to the lapse and shift, $N$ and $N^i$, generates the equations of motion (constraints) that need to be satisfied. To see this, we expand the Einstein-Hilbert action in terms of the ADM variables, 
\beq
{\cal L}_{\rm EH} = \frac{\Mp^2}{2} \sqrt{-g} R = \frac{\Mp^2}{2} a^3(t)[ N R^{(3)} + \frac{1}{N} (E^{ij} E_{ij} - (E^{i}_{i})^2) ]  \ ,
\eeq
where
\beq
E_{ij} = \frac{1}{2} [ \dot g_{ij} - \hat \nabla_i g_{j k} N^k - \hat \nabla_j g_{i k} N^k ] = \frac{1}{2} [ 2 H a^2 \delta_{ij} - a^2 \partial_i N_j - a^2 \partial_j N_i ] \ ,
\eeq
$E^{ij}=g^{im} g^{jm} E_{nm}$, $g_{ij} =\delta_{ij} a^2$, and $N_i$ is lowered with $\delta_{ij}$.  We used the choice of gauge to simplify the expression, since we have chosen a flat slicing such that there is no curvature on the slice (i.e. the covariant derivatives are ordinary derivatives) and set $R^{(3)} = 0$.

Let's work to further simplify the gravitational action.  We will write out the extrinsic curvature terms,
\bea
E_{ij} E^{ij} &=& \frac{1}{4} [ 2 H \delta_{ij} - \partial_i N_j - \partial_j N_i ][ 2 H \delta^{ij} - \partial^i N^j - \partial^j N^i ] \\
&=& 3 H^2 - 2 H \partial_i N^i + \frac{1}{2}( \partial_i N_j \partial^i N^j +  \partial_i N_j \partial^j N^i ) 
\eea
and
\beq
E_i^i =  3 H - \partial_i N^i \ .
\eeq
Putting this all together, the Einstein action is given by
\bea
{\cal L}_{\rm EH} &=& \frac{\Mp^2}{2} \frac{a^3(t)}{N} [ -6 H^2 +4 H \partial_i N^i + \frac{1}{2}( \partial_i N_j \partial^i N^j +  \partial_i N_j \partial^j N^i )  - (\partial_i N^i )^2  ] \ .
\eea
We will work with a  small $c_s$ model, so that
\bea
{S}_{\pi} &=& \int d^4 x a^3 N [ - \Mp^2 (3 H^2 + \dot H) + \Mp^2 \dot H (- \frac{1}{N^2} (1+\dot \pi - N^i \partial_i \pi)^2 + a^{-2} \partial_i \pi \partial^i \pi)  \nonumber \\ && \qquad + \frac{M^4_2}{2}  (- \frac{1}{N^2} (1+\dot \pi - N^i \partial_i \pi)^2 + a^{-2} \partial_i \pi \partial^i \pi +1)^2 \ .
\eea
Neither $N$ nor $N^i$ has a kinetic term and thus they are both constrained variables. The equations of motion at ${\cal O}(\pi^0)$ are equivalent to Einstein's equations for the background and are satisfied for $N=1$ and $N^i = 0$. At linear order, ${\cal O}(\pi)$, we can solve $N=1 + \delta N$ to find
\beq
\delta N = \epsilon H \pi \qquad \partial^i N_i = - \frac{H \epsilon \dot \pi}{c_s^2} \ .
\eeq
The complete derivation is provided in Appendix~\ref{app:ADM}.

We notice that the solution for $N^i$ scales as $\epsilon / c_s^2$. A priori, it is possible to have $\epsilon \ll 1$ as required for inflation, but still having $\epsilon >  c_s^2$ such that $N^i$ is not manifestly small. This might suggest that the coupling to gravity may not be negligible when $c_s^2 \ll 1$. However, when we plug these constraints back into the action, we find the Lagrangian at quadratic order in $\pi$, ${\cal L}_{\pi,2}$, becomes
\beq
{\cal L}_{\pi, 2} = -  \frac{\Mp^2 \dot H}{c_s^2} (\dot \pi^2 - c_s^2 \frac{\partial_i \pi \partial^i \pi}{a^2} + 3 H^2 \epsilon \pi^2 ) \ .
\eeq
The factor of $c_s^{-2}$ simply combines with the normalization of the kinetic term so that the gravitational correction is suppressed by $\epsilon \ll 1$. 

What remains to be answered is whether all gravitational corrections are similarly suppressed by $\epsilon$. To do so, we will explore the structure of the gravitational corrections to the action and the physical origin of the $c_s^{-2}$ enhanced contribution to $N^i$. Interestingly, a similar investigation of the structure of the ADM constraints played a non-trivial role in defining the soft behavior of $\zeta$~\cite{Weinberg:2003sw,Creminelli:2012ed,Hinterbichler:2012nm,Hinterbichler:2013dpa,Pajer:2013ana}. It is therefore fitting that a complete understanding of the gravitational corrections to the action of $\pi$ also requires a detailed examination of the very same constraints.

\section{Goldstone Boson Equivalence}\label{sec:cs}

In the previous section, we saw that the decoupling limit suggests the Goldstone boson action, on a fixed de Sitter metric, should describe the scalar fluctuations during inflation up to slow roll corrections. In this section, we will demonstrate that gravitational corrections are controlled by an expansion in $\epsilon$. We will first derive expression for the leading $\epsilon$ correction, valid to all-orders in $\pi$, and then show how it generalizes to higher-orders in $\epsilon$.

\subsection{Linearized Gravity and the Stress Tensor}

In this section, we wish to streamline the derivation of the contributions to the action for $\pi$ from the constraints imposed by $N$ and $N^i$. A priori, the coupling of gravity to the theory of inflation involves a mess of parameters that characterize the inflationary model. The central observation that makes the EFT of inflation powerful is that there is a hierarchy of energy scales
\beq
\Mpl^4 \gg \Mpl^2 H^2 \gg f_\pi^4 \gg H^4 \gg \dot H^2 \ .
\eeq
Most notably, since the Planck scale is the largest scale, we should expect that the corrections to the gravitational action from the inflationary dynamics are small. However, our challenge is writing the corrections in such a way that this is manifest.

The key observation is that the coupling between $\pi$ and the metric, at linear order in $N$, $N^i$, but all-orders of the $\pi$, is determined by the $\pi$ stress tensor, $T_{\mu \nu}$, on a fixed de Sitter background. Specifically, if we write the full action as  
\beq
S = S_{\rm EH} + S_\pi \ ,
\eeq
then the sources for $\delta N$ and $N^i$ are just the linear couplings to the metric,
\beq 
\frac{\delta S_\pi}{\delta N}|_{N=1,N^i=0} = - a^3 T_{00} 
 \qquad \frac{\delta S_\pi}{\delta N^j}|_{N=1,N^i=0} =- \frac{a^3}{2} \left( T_{0j} + T_{j0} \right) \ ,
\eeq
where $T_{\mu \nu}$ is computed on a fixed de Sitter background (see Appendix~\ref{app:stress} for details and derivations of the properties of the stress tensor for the inflaton).

Meanwhile, the quadratic terms in the metric are given by the Einstein-Hilbert action
\beq
{\cal L}_{\rm EH}^{(2)} =\Mp^2 \left( 3 H^2(\delta N - \delta N^2) -2 H \delta N \partial_i N^i +\tfrac{1}{4}( \partial_i N_j \partial^i N^j +  \partial_i N_j \partial^j N^i )  -\tfrac{1}{2} (\partial_i N^i )^2 \right) 
\eeq
Differentiating, the constraint equations for $\delta N$ and $N^i$ become
\bea
-6 \Mpl^2 H^2 \delta N- 2 \Mp^2 H \partial^2 \chi&=& \delta T_{00} \ , \\
 2 \Mpl^2 H \partial_j \delta N  - \frac{\Mp^2}{2} (\partial^2 \Psi_j) &=&  T_{0 j} \ ,  
\eea
 where we will split $N^i = \partial^i \chi + \Psi^i$ with $\partial_i \Psi^i = 0$, and defined $\delta T_{00} = T_{00} - 3\Mpl^2 H^2(t)$ so that $\delta T_{00} = T_{0j} = N^i=0$ and $N=1$ when $\pi=0$. These equations can be solved straightforwardly to give $\Psi_i=0$,
 \beq\label{eq:general_lin}
\delta N = \frac{\partial^j}{\partial^2}\frac{ \delta T_{0 j}}{2 \Mpl^2 H} \ , \qquad \partial^2 \chi = -\frac{1}{2 \Mpl^2 H} \left(\delta T_{00}+3 H \frac{\partial^j}{\partial^2} T_{0j} \right) \ .
\eeq 
At linear order in $\pi$, using 
\beq
\delta T_{00}= - 2 \frac{\Mp^2 \dot H}{c_s^{2}} \dot \pi +6 \Mpl^2 H \dot H \pi + {\cal O}(\pi^2) \qquad
\delta T_{0j} = - 2 \Mp^2 \dot H \partial_j \pi + {\cal O}(\pi^2) \ ,
\eeq
we find that 
\beq
\delta N  =  H \epsilon \pi + {\cal O}(\pi^2) \qquad \partial^2 \chi   = -\frac{H \epsilon}{c_s^2 } \dot\pi + {\cal O}(\pi^2)
\eeq
as before. 

The benefit of this description is that we are not limited to a particular order in $\pi$. Instead, we can substitute our all-orders solution in terms of $T_{\mu \nu}$ directly into the action to get
\beq
\delta {\cal L}_{\pi} = \frac{3}{4 \Mpl^2} \left(\frac{\partial^j}{\partial^2} \delta T_{0j} \right)^2 + \frac{1}{2 \Mpl^2 H} T_{0j} \frac{\partial^j}{\partial^2} \left(\delta T_{00}+ 3 H \frac{\partial^j}{\partial^2} T_{0j} \right)
\eeq
At finite momentum $\partial^j \to - i k^j \neq 0$, we can ``integrate by parts" (use momentum conservation) to simplify this expression
\beq
\delta {\cal L}_{\pi} = -\frac{3}{4 \Mpl^2} \left(\frac{\partial^j}{\partial^2} \delta T_{0j} \right)^2 + \frac{1}{2 \Mpl^2 H} T_{0j} \frac{\partial^j}{\partial^2} \delta T_{00}
\eeq
While this procedure has potential issues at zero momentum, it is sufficient for most tree-level cosmological correlator calculations.

The above result is the general leading gravitational correction to the Goldstone boson action. Yet, it does not immediately show how these corrections compare to the terms at the same order in the decoupled Goldstone boson action. To determine the size of the gravitational terms, we determine the stress tensor using
\beq
T_{\mu \nu}=-2 \frac{\partial \mathcal{L}_{\pi }}{\partial g^{\mu \nu}}+g_{\mu \nu} \mathcal{L}_{\pi} \ ,
\eeq
in the limit of the EFT of inflation where we include only the $M_n^4$ terms, namely Equation~(\ref{eq:Mn_action}). The resulting Goldstone boson stress tensor is then given by 
\beq
T_{\mu \nu} = - 2 \partial_\mu U \partial_\nu U \sum_{n=0}^{\infty} \frac{M^4_{n+1}(U)}{n!} (\partial_\mu U \partial^\mu U+1)^n  + g_{\mu \nu} \sum_{n=0}^{\infty}  \frac{M^4_n(U)}{n!}  (\partial_\mu U \partial^\mu U+1)^n \ . 
\eeq
Taking $g_{\mu\nu}$ to be the fixed background metric, we can derive the leading gravitational correction to the action at all orders in $\pi$. First we need to remove the background, to define $\delta T_{\mu \nu} = T_{\mu \nu} - T_{\mu \nu}|_{\pi=0}$,
\begin{align}
\delta T_{\mu \nu} =&- 2 \left( M_1^4(U) \partial_\mu U \partial_\nu U - M^4_1(t) \delta_{\mu 0 } \delta_{\nu 0} \right) + g_{\mu \nu} \left( M_0^4(U) - M^4_0(t) \right) \\
&\sum_{n=1}^{\infty}\frac{1}{n!}(\partial_\mu U \partial^\mu U+1)^n \left(   - 2 M^4_{n+1}(U) \partial_\mu U \partial_\nu U  + g_{\mu \nu} M^4_n(U) \right) \ . 
\end{align}
Fortunately, since $\partial^i T_{\mu \nu}(\pi=0) =0$ and $T_{0j}(\pi=0) =0$, we can simplify the expression for the contribution to the action, provided we evaluate $\partial^j /\partial^2 T_{00} = \partial^j /\partial^2 \delta T_{00}$. With these assumptions, we can write
\beq
\begin{aligned}
\delta {\cal L}_{\pi} =&- \frac{3}{ \Mpl^2} \sum_{m,n=0}^\infty  \frac{ M_{m+1}^4 M^4_{n+1}}{n! m! } \\
&\times\left[ \frac{\partial^j}{\partial^2} \left( \partial_j U \partial_0 U \left(\partial_\mu U \partial^\mu U + 1)^m \right) \right) \frac{\partial^k}{\partial^2} \left( \partial_k U \partial^0 U  \left(\partial_\mu U \partial^\mu U + 1)^N \right) \right) \right]  \\
& -\frac{1}{\Mpl^2 H} \sum_{m,n=0}^\infty  \frac{M_{m+1}^4 }{ m! n!} \bigg[  \partial_j U \partial_0 U \left(\partial_\mu U \partial^\mu U + 1)^m \right) \\
&\times\left(\frac{\partial^j}{\partial^2} \left(\partial_\mu U \partial^\mu U + 1 \right)^n \left(- 2 M^4_{n+1}(U) \partial_0 U \partial_0 U  - M^4_n(U) \right)  \right) \bigg] \ .
\end{aligned}
\eeq
As written, it is hardly clear that these corrections to the action are small compared the terms already present in the decoupling limit. In fact, for sufficiently weak interactions, we know that gravitational corrections will be important, as these are responsible for non-Gaussianity in slow-roll inflation\cite{Maldacena:2002vr}. 

In order to quantify the importance of these new terms, it is useful to define~\cite{Baumann:2015nta}
\beq
M_n^4 \equiv c_n \frac{f_\pi^4}{c_s^{2 n-1}} \ .
\eeq
This definition is such that when $c_s \ll 1$, all the higher order interactions are controlled by a single UV scale $\Lambda = f_\pi c_s$ for $c_n = {\cal O}(1)$. This scaling is also expected by naturalness of the coefficients~\cite{Senatore:2010jy,Baumann:2011su} and consistency of scattering in the flat space limit~\cite{Baumann:2014cja,Baumann:2015nta,Baumann:2019ghk,Grall:2020ibl}. Now, using
\beq
\partial_j U \partial_0 U  = {\cal O}(\pi) \qquad \partial_\mu U \partial^\mu U + 1 = {\cal O}(\pi) \ ,
\eeq
we can compare the contribution at ${\cal O}(\pi^N)$ between the the decoupled action,
\beq
{\cal L}_{\pi, N} \supset  \frac{M_{N}^4}{N!} (-2 \dot \pi)^N \ ,
\eeq
and the leading gravitational correction at the same order, 
\bea
\delta {\cal L}_{\pi, N} &\supset& \frac{2 M_{m+1}^4 M^4_{n+1}}{n! m! \Mpl^2 H} \partial_j \pi (-2 \dot \pi)^m \frac{\partial^j}{\partial^2} (-2 \dot \pi)^{n}  \ ,
\eea
where $N=n+m+1$, and ${\cal L}_{\pi,N}$ denotes a term at order ${\cal O}(\pi^N)$. Comparing the sizes of these terms, we find
\bea
 \left| \frac{\delta {\cal L}_{\pi,N}}{{\cal L}_{\pi,N} }\right| &\approx&  \frac{N!}{4 n! m! } \frac{2 M_{m+1}^4 M^4_{n+1}}{M_N^4 \Mpl^2 H^2}\left( \frac{K_j}{K^2 } \frac{H K^j \pi}{\dot \pi} \right)   \\
 &\approx&  \frac{N!}{n! m! } \frac{2 c_{m+1} c_{n} }{c_N} \frac{f_\pi^4}{c_s \Mpl^2 H^2} = \frac{ N!}{ n! m! } \frac{2 c_{m+1} c_{n} }{c_N} \epsilon  \ , \label{eq:estimate}
\eea
where $\vec K$ is some linear combination of momenta, $\dot \pi \sim \omega \pi$, and we used $\omega = c_s k = H$ to estimate this contribution at horizon crossing with $K \sim K' \sim k$.

Recall that for all $N$, $c_N ={\cal O}(1)$ is the expected size of the higher order terms in the decoupled action such that the gravitational terms are necessarily suppressed by $\epsilon \ll 1$. However, for very weakly coupled models, like those that occur in slow roll, $c_N \ll 1$ for $N> 1$ and therefore, there can be competition between terms. Finally, the enhancement of this ratio by the factor of $N!$ is the familiar fact that our perturbative expressions are only a asymptotic series.

Our estimate of the size of the gravitational correction does not include all terms at order $\pi^N$, as we have only included the leading power of $\pi$ for each $M^4_{n}$. Of course, each coupling is associated with several power of $\pi$ due to the nonlinearly realized symmetry. These higher-order terms are additionally suppressed by powers $H^2 /f_\pi^2$ (i.e.~the amplitude of the scalar power spectrum) in the contributions to correlators, although they may appear to be significant prior to canonical normalization of $\pi$ or using $\omega = c_s k \ll k$ when $c_s \ll 1$. One can check that the definition of $M_n$ in terms of $c_n = {\cal O}(1)$ ensures the sub-leading powers of $\pi$ do not lead to parametrically enhanced interactions.

In order to illustrate the subtlety in comparing the sizes of the gravitational terms, let us consider the cubic gravitational correction 
\beq
 \delta {\cal L}_{\pi,3} \supset \frac{4 M_{2}^8}{\Mpl^2 H} \bigg[  \partial_j \pi \dot \pi   \frac{\partial^j}{\partial^2}\dot \pi \bigg] \approx \frac{f_\pi^4}{c_s^5} \frac{f_\pi^4}{\Mpl^2 H^2 c_s} \dot \pi^3  =  \epsilon \frac{f_\pi^4}{c_s^5} \dot \pi^3 \ ,
\eeq
where we estimated $\partial_j \pi \to k_j  \pi \approx \dot \pi/c_s$ at horizon crossing. To reach the comparison in Equation~(\ref{eq:estimate}), we look at the term in the decoupled action proportional to $M_3^4$ ($n=m=1$ and $N=3$) , 
\beq
{\cal L}_{\pi,3}\supset M_3^4 \dot \pi^3 =c_3 \frac{f_\pi^4}{c_s^5 }\dot \pi^3 \ .
\eeq
As expected, this correction is clearly subleading by a power of $\epsilon$. Furthermore, the conclusion is not limited to $c_3$ alone. Suppose we set $c_3=0$ while holding $c_s \ll 1$. In this case, we still have to compare our gravitational corrections to the cubic terms generated by $M_2^4$. Concretely, we have the additional term  
\beq
{\cal L}_{\pi,3}\supset M_2^4 \dot \pi \partial_i \pi \partial^i \pi = \frac{f_\pi^4}{c_s^3 }\dot \pi \partial_i \pi \partial^i \pi \approx \frac{f_\pi^4}{c_s^5} \dot \pi^3 \ ,
\eeq
which scales the same as $c_3 = {\cal O}(1)$ after using $\partial_i \pi \partial^i \pi \sim c_s^{-2} \dot \pi^2$. In this sense, we see that the gravitational contributions are suppressed, even though they may appear with additional larger powers of $c_s^{-2}$.

\subsection{Higher-Order Gravitational Corrections}

Assuming a controlled expansion in $\epsilon$, we have shown that the first correction is sub-leading to the decoupled Goldstone boson action. Here we will demonstrate how the expansion to higher-orders is organized and show that the explicit calculation of the second order correction is consistent with this behavior.

\subsection*{Scaling with Speed of Sound }

When showing the linear correction from coupling to gravity is ${\cal O}(\epsilon)$, we implicitly ignored higher order terms in $\delta N$ in the action for $\pi$ assuming they were additionally suppressed by $\epsilon$. For this to be true, we require that gravitational terms in the $\pi$ action are smaller than contributions of the same order in $\delta N$ from the Einstein-Hilbert terms,~e.g.
\beq\label{eq:M2ineq}
M_2^4 \delta N^2  \ll -3 \Mpl^2 H^2 \delta N^2 -2 \Mpl^2 \delta N \partial_i N^i \ .
\eeq
If $c_s^2 < \epsilon$ then $M_2^4 \approx  \Mpl^2 \dot H/c_s^2 > \Mpl^2 H^2$ and it is not obvious that Equation~(\ref{eq:M2ineq}) is valid. Therefore, we need a better understanding of how these ``higher-order" corrections are generated to be sure that the expansion is under control. Just from examining the sizes of coefficients seems to be insufficient.

In order to make sense of factors of $c_s$, it is important to take into account the different in units of $N$ and $N^i$ or, equivalently, $T_{00}$ and $T_{0i}$. From the line element, 
\beq
ds^2 = - N^2 dt^2 + a(t)^2\delta_{ij} (d x^i + N^i dt) (d x^j+ N^j dt) \ ,
\eeq
we see $N^i$ and $N$ must have units (where we denote the units of $X$ as $[X]$) such that
\beq
[N_i][dt][dx^i] = [N]^2 [dt]^2 \to [N^i] = [N]^2 \frac{[dt]}{[dx]}
\eeq
Since we are neglecting gravitational waves, the only velocity in the problem is $c_s$ and therefore, we should expect $N^i \propto c_s^{-1}$. From our explicit solution, using $\partial^i \to k^i \approx \omega/c_s$, we see 
\beq
\partial^i N_i = - \frac{H \epsilon \dot \pi}{c_s^2} \to N^i = H\frac{k_i}{k^2} \frac{\epsilon \omega \pi}{c_s^2} \approx \frac{H \epsilon}{c_s} \pi \ ,
\eeq
as expected from units. As a result, we have
\beq
\Mpl^2 \delta N \partial_i N^i \approx \frac{\Mpl^2 H^2}{c_s^2} \delta N^2 \approx \frac{M_2^4}{\epsilon} \delta N^2 \ .
\eeq
Therefore, we can recognize that then dominant gravitational term in the Lagrangian is $\Mpl^2  \delta N  \partial_i N^i$ which is larger than any contribution in the $\pi$ Lagrangian.

When we substitute this back into the action, it is also important that $N^i$ appears in the form
\beq
g^{\mu \nu} \partial_\mu(t+\pi)\partial_{\nu}(t+\pi) \supset \partial_i \pi \partial^i \pi - 2 N^i \partial_i \pi \ .
\eeq
Therefore, the gravitational corrections scale as
\beq\label{eq:scaling_Ni}
\frac{|N^i|}{|\partial_i \pi|} \approx \epsilon \ , 
\eeq
with no additional $c_s$ factors.

\subsection*{Structure of the Expansion}

From the linearized coupling of gravity to the inflaton, we were able to calculate the leading correction in $\epsilon$ expansion, while keeping all orders in the Goldstone boson $\pi$. Now we want to demonstrate that this procedure can be applied order by order in the slow-roll parameter
\beq
\epsilon  \equiv -\frac{\dot H}{H^2} =  \frac{f_\pi^4}{c_s\Mpl^2 H^2}  \ll 1 \ ,
\eeq
where some terms are additionally suppressed by powers of $c_s \leq 1$. Here we will solve for
\bea
\delta N &=& \delta N^{(1)} + \delta N^{(2)} +\ldots \\
N^i &=& N^{i,(1)} + N^{i,(2)} +\ldots 
\eea
where $N=1+\delta N$ and $\delta N^{(n)}, N^{i,(n)} = {\cal O}(\epsilon^n)$.

What simplified this process at ${\cal O}(\epsilon)$ is that the key coupling of $N$ and $N^i$ to $\pi$ was linear and therefore depends only on the stress tensor in the fixed background space. At ${\cal O}(\epsilon^{n>1})$, we must account for nonlinear (model-dependent) metric couplings in the action for $\pi$.

Both the Einstein-Hilbert action and the action for $\pi$ contain higher orders in $\delta N$ and $N^i$. The expansion in $\epsilon$ arises because of two features of these corrections. First, the Einstein-Hilbert term is proportional to $\Mpl^2 H^2$ and starts at quadratic order in $\delta N$, $N^i$. Second, the action for $\pi$, and metric components therein, are proportional to $f_\pi^4 = \epsilon \Mpl^2 H^2/c_s^2$ and start at linear order in $\delta N$, $N^i$. As a result, at $(m+1)$-th order in $\epsilon$, the constraint equations will take the form
\begin{align}
-6  \delta N^{(m+1)}- 2  H^{-1} \partial^2 \chi^{(m+1)}  &= \frac{\epsilon}{c_s^2} \left( \delta N^{(m)} F (\pi)+  N^{k,(m)} G_k (\pi) +\ldots \right) \\
 2 H^{-1} \partial_j \delta N^{(m+1)}  - \frac{1}{2 H^2} (\partial^2 \Psi^{(m+1)}_j) &=  \epsilon \left(  \delta N^{(m)} H_j (\pi) + N^{k,(m)} J_{j k}(\pi)  + \ldots\right)
\end{align}
Here the $\ldots$ includes higher orders terms in the lapse and shift that are the same order in $\epsilon$ as $\delta N^{(m)}$ or $N_i^{(m)}$, such as $\delta N^{(m-p)} \delta N^{(p)}$ through $(\delta N^{(1)})^m$. These non-linear terms are the reason that the solution at higher orders is non-trivial, but for the purpose of overall scaling, it only matters that they are of the same order as $N^{(m)}$ or $N^{k,(m)}$. The function $F(\pi), G_k(\pi), H(\pi)$,and $J_k(\pi)$ are all orders in $\pi$ but linear in $M^4_n /\epsilon$. For the purpose of our power counting, these functions are ${\cal O}(1)$. 

Expanding the solutions for $\delta N$ and $N^i$ by counting powers of $\epsilon$ is unambiguous, and ultimately explains the small size of the higher-order gravitational corrections. However, in writing the equation at order $(m+1)$, we also included an explicit factor of $c_s^{-2}$. This factor was introduced to account for the difference of units between the two equations. We will keep track of these factors of $c_s$ while ignoring of factors of $c_s$ associated with $M_n^4$ or $\pi$. While somewhat ambiguous, this tells us the {\it naive} scaling with $c_s$ that we should expect at order $m$ that follows entirely from the units of $N$ and $N^i$. As shown in Equation~(\ref{eq:scaling_Ni}), these factors should cancel when comparing dimensionless quantities. Additional factors of $c_s$ may arise through functions of $\pi$, e.g.~$F(\pi)$, but these factors were already present in the decoupled action and thus do not enhance the size of the gravitational corrections. We will confirm this behavior with our $m=1$ and $m=2$ solutions.

Now suppose we assume the scaling at ${\cal O}(\epsilon^m)$ takes the form,
\beq
\delta N^{(m)} = {\cal O}\left(\frac{\epsilon^{m}}{c_s^{2m-2}}\right) \qquad N_i^{(m)} = {\cal O}\left(\frac{\epsilon^{m}}{c_s^{2m}} \right) \  ,
\eeq
where the factors of $c_s$ are again the naive (explicit) behavior. We already know this scaling holds for $m=1$, and we will demonstrate it also holds for $m=2$. To satisfy this equation, the solutions must scale as
\beq
\delta N^{(m+1)} = {\cal O}\left(\frac{\epsilon^{m+1}}{c_s^{2m}}\right) \qquad N_i^{(m+1)} = {\cal O}\left(\frac{\epsilon^{m+1}}{c_s^{2m+2}} \right) \  .
\eeq
We see that the contribution from $N_i^{(m)}$ is responsible for generating the largest explicit inverse powers of $c_s$ in both $\delta N^{(m+1)}$ and $N_i^{(m+1)}$. Since these factors are a direct consequence of the units of $N_i$, starting at $m=1$, each of these terms should be associated with factors of $(\partial \pi)^m$, in order for the units to match. From  Equation~(\ref{eq:scaling_Ni}), we should anticipate that these higher-order terms will be suppressed by powers of $\epsilon$ relative to the corresponding term in $S_\pi$

\subsubsection*{Explicit Result at Second Order} 

Now we would like to verify at ${\cal O}(\epsilon^2)$ that our gravitational corrections are consistent with our expected scaling. Related results were derived in slow-roll ($c_s=1$) in~\cite{Seery:2006vu,Pajer:2016ieg,Bonifacio:2022vwa} for the purpose of calculating the trispectrum, but our focus here is illustrating the expansion in $\epsilon$ in a concrete example. Related calculations of the trispectrum with $c_s \ll 1$ have (typically) assumed these corrections are subdominant~\cite{Chen:2006dfn,Chen:2009bc}, as demonstrated here.

We can define the constraint equations for $N$ and $N^i$ non-perturbatively using the variation of the action for the Goldstone, 
\begin{equation}
	\begin{aligned}
	\frac{\delta S_{\pi}}{\delta N} &= -N^{-2} a^3 (T_{00} - 2 T_{0i} N^i - N^i N^j T_{ij})\\
	\frac{\delta S_{\pi}}{\delta N^j} &= -\frac{a^3}{N} (T_{0j} - N^i T_{ij}) \ ,
	\end{aligned} \ ,
\end{equation}
and the Einstein Hilbert term
\begin{equation}
	\frac{\delta S_G}{\delta N} = -\Mpl^2 a^3\frac{1}{2N^2} (E_{ij} E^{ij} - E^2) \hspace{1cm} \frac{\delta S_G}{\delta N^j} = \Mpl^2 a^3 \nabla_i \bigg[E^i_{\,\,j} - \delta^i_{\,\,j} E \bigg] \ .
\end{equation}
These are, of course, just a rephrasing of the Einstein's field equations. However, after fixing to the $\pi$-gauge, actually solving the equations at a fixed order will be easier in this language.

Now we want to expand both sides to order $\epsilon^2$, treating the scaling with $c_s$ as ${\cal O}(1)$. Details can be found in Appendix~\ref{app:esp2}. We have already solved for ${\cal O}(\epsilon)$ term, so after removing the first order terms from the equations. Expanding the stress tensor in $\epsilon$, via
\beq
T_{\mu \nu} = \sum_{n=0}^\infty T^{(n)}_{\mu \nu} \qquad T^{(n)}_{\mu \nu} = {\cal O}\left( \epsilon^n \right) \qquad M_n^4,\delta N, N^i ={\cal O}(\epsilon) \qquad \pi = {\cal O}(1) \ ,
\eeq
The equations of motion for the lapse and shift at second order become
\begin{align}
	\Mpl^2  \bigg[- 6 H^2 \delta N^{(2)} &-2 H \partial_i N^{i,(2)} + 4 H \delta N \partial_i N^i+9 H^2 (\delta N)^2  \nonumber\\
	   &-\frac{1}{4} (\partial^i N_j \partial_i N^j + \partial^i N_j \partial_j N^i - 2 \partial_i N^i \partial_j N^j) \bigg]  \\
	   &= 3 \overline{T}^{(0)}_{00} (\delta N)^2 - 2 \delta N \delta T_{00}^{(1)} + \delta T_{00}^{(2)} - 2 N^i T_{0i}^{(1)} + N^i N^j \overline{T}^{(0)}_{ij} \nonumber  \ ,
    \end{align}
and
 \begin{align}
 \Mpl^2  \bigg[ - 2H \partial_j \delta N^{(2)} &+ \frac{1}{2}(\partial^2 N_j^{(2)} - \partial^j \partial_i N^{i,(2)}) - \frac{1}{2}(\delta N) ( \partial^2 N_j - \partial^j \partial_i N^i) \nonumber \\
	& - 2 H \delta N \, \partial_j (\delta N) - \frac{1}{2}\partial_i (\delta N) (\partial^i N_j + \partial^j N_i - 2 \delta_{ij} \partial^k N_k) \bigg] \\
	 & = T_{0j}^{(2)} - \delta N T_{0j}^{(1)} - N^i T_{ij}^{(1)} + \delta N N^i \overline{T}^{(0)}_{ij} \ . \nonumber	 
\end{align}
Here we also used the shorthand $\delta N, N_i= \delta N^{(1)}, N_i^{(1)}$ unless otherwise specified. This equations specify the all-orders solutions in $\pi$ at ${\cal O}(\epsilon^2)$. However, we will be more interested in the explicit solutions in terms of $\pi$ for comparison to the expected scaling behavior.

We can solve the equations explicitly for $M_{0,1,2}^4 \neq 0$ keeping up to ${\cal O}(\pi^2)$ to find
\begin{align}
	\delta N^{(2)} =& \frac{\partial_j}{\partial^2} \bigg [ \frac{\epsilon^2}{c_s^2}(1 - c_s^2) H^2 \pi \partial_j \pi - \frac{\epsilon^2}{c_s^2} \frac{H}{4} \left (\dot{\pi}\partial_j \pi - \partial_i \pi \frac{\partial_i \partial_j}{\partial^2}  \dot{\pi} \right ) \bigg]
\end{align}
and
\begin{align}
	\partial^2 \chi^{(2)} =& -3 H \delta N^{(2)} + \frac{\epsilon^2}{c_s^2}H^2 \pi - \frac{\epsilon^2 H^3}{4 c_s^4}\left ( \left ( \frac{\partial_i \partial_j}{\partial^2} \dot{\pi} \right )^2 - \dot{\pi}^2\right ) + 3 \frac{\epsilon^2}{c_s^2} H^2 \dot{\pi} \pi \nonumber \\
	 & + \frac{\epsilon^2}{c_s^4} H^2 \frac{\partial_i}{\partial^2} \dot{\pi} \partial^i \pi  - 3\epsilon^2 H^2 \dot{\pi} \pi  + \frac{21}{2} \epsilon^2 H^3 \pi^2 \ .
  \end{align}
We see that $\delta N^{(2)} = {\cal O}(\epsilon^2 / c_s^2)$ and $\partial^2 \chi^{(2)} = {\cal O}(\epsilon^2/c_s^4)$ as expected from our general scaling argument.

\section{Generalizations}\label{sec:generalizations}

Like any effective theory, the EFT of inflation includes higher-derivative terms. For example, if we define
\beq
   \Big[\nabla_\mu \nabla_\nu U  \Big]  \equiv  \Big(\nabla_\mu \nabla_\nu U - H(g_{\mu \nu} + \nabla_{\mu} U \nabla_{\nu} U) \Big) \ ,
\eeq
then we can consider adding terms to the action of the form
\beq
S_\pi  \supset \int d^4 x \sqrt{-g} \sum_{n \ge 2} \bar M_n^{4-n}(U) \Big[ g^{\mu \nu} \nabla_\mu \nabla_\nu U  \Big]^n\ .
\eeq
In most models, these corrections are subleading to contributions from $M_n^4$. As a result, their contributions to the mixing with gravity will be suppressed by powers of $\epsilon$ for the same reason that the contribution from $M_n^4$ are suppressed. 

In some models, higher derivative terms alter the dispersion relation, so that $\omega  = f(k)$ where the function $f(k)$ is non-linear at horizon crossing, $\omega  = H$. The power counting argument in Section~\ref{sec:cs} crucially uses $\omega = c_s k$ to estimate the size of the gravitational terms and thus could be significantly altered by a change to the dispersion relation. Nevertheless, at linear order in $\delta N$ and $N^i$, the result from Equation~(\ref{eq:general_lin}) remains correct
\beq
\delta N = -\frac{\partial^j}{\partial^2}\frac{ \delta T^{0}_j}{2 \Mpl^2 H} \qquad \partial^2 \chi = -\frac{1}{2 \Mpl^2 H} \left(\delta T^{00}- 3 H \frac{\partial^j}{\partial^2} T^0_j \right) \ .
\eeq
Furthermore, if our kinetic term is $\dot \pi^2$,  at linear-order in $\pi$ the stress tensor is given by~\cite{Baumann:2011ws}
\beq
T^0_0 = \bar f^4 \dot \pi \qquad T^{i}_0 = \partial^i {\cal O} \ ,
\eeq
where $\bar f^4$ is a constant with units of energy density, and ${\cal O}$ is an operator linear in $\pi$ so that $\bar f^4 \ddot \pi  - \partial_i \partial^i {\cal O} =0$ gives the quadratic equations of motion. The relationship between $\bar f$ and $f_\pi$, the energy scale, for models with non-linear dispersion relations requires more care than the $c_s$ case and is discussed in detail in~\cite{Baumann:2011ws}.

The case of ghost inflation~\cite{Arkani-Hamed:2003juy} provides a concrete example of this behavior. It can be described by the limit $\Mpl^2 \dot H \to 0$ with $M_2^4 \neq 0$ and $\bar M_2^2 \neq 0$. From this action, we again find the stress tensor at linear order in $\pi$, 
\beq
T_{00} = M_2^4 \dot \pi \qquad T^{0i} = - \bar M_2^2 \nabla_i \nabla^2 \pi
\eeq
If we define $\rho^2 = M_2^4 /\bar M_2^2$ then the dispersion relation is $\omega^2 =k^4/\rho^2$ as required by conservation of the stress tensor. We can then find the solution to the constraints
\beq
\delta N =\frac{M_2^4}{2 \Mpl^2 H^2} \frac{H\partial^2 \pi}{\rho^2} \qquad \partial^2 \chi = -\frac{M_2^4}{2 \Mpl^2 H} \dot \pi -\frac{3 M_2^4}{2\Mpl^2} \frac{\partial^2 \pi}{\rho^2}  \ .
\eeq
Notice that at horizon crossing, $\omega =k^2/\rho = H$, we can write the phase velocity 
\beq
c^2_p =\omega^2 / k^2 = \frac{H}{\rho}
\eeq
then
\beq
\delta N =\frac{M_2^4}{2 \Mpl^2 H^2} \frac{H\partial^2 \pi}{\rho^2} \approx \frac{M_2^4}{2 \Mpl^2 H^2} H c_p^2 \pi 
\eeq
So that the solution for $\delta N$ is suppressed by a factor of the small phase velocity $c_p^2 \ll 1$ relative to $N^i$, just like we found in the case of small $c_s$. Furthermore, since $k = c_p \omega$ (by definition), the same dimensional analysis argument suggests that $N^i$ will be suppressed relative to $\partial_i \pi$.

The structure of the constraints is a natural generalization of the small $c_s$ EFT. However, we saw that for all the higher order corrections to be suppressed, it was important that $M_n^4 \propto c_s^{-2n +1}$ and that $f_\pi^4/(c_s \Mpl^2 H^2)= \epsilon \ll 1$. In principle, a definitive generalization of the equivalence theorem to arbitrary dispersion relation requires a similar understanding of natural size of the higher-dimension operators. There is no obvious obstacle to deriving the correction scaling on a case by case basis, but it would be preferable to establish a model-independent explanation for the required structure.

\section{Conclusions}\label{sec:conclusions}

Cosmological correlators are a topic of broad interest, both as a formal arena for understanding the dynamics of accelerating spacetimes~\cite{Flauger:2022hie} and as an observable probe of the very early universe~\cite{Achucarro:2022qrl}. While the rules for calculating the predictions of a given model are well known~\cite{Weinberg:2005vy}, it is desirable to understand how different physical effects manifest themselves in the final result. Applying such knowledge in reverse allows one to bootstrap the form of the full correlators from the requirement that it reproduces known physics in specific limits~\cite{Baumann:2022jpr}.

The decoupling of gravity is one tool that has been enormously useful in classifying the space of large non-Gaussian signals~\cite{Cheung:2007st}. It ensures that the dominant contributions to many higher-point correlators are the self-interactions of the inflaton, as described by the Goldstone boson $\pi$, rather than interactions mediated by gravity. While it is intuitively clear this decoupling is controlled in the $\Mpl \to \infty$ limit, a derivation of the equivalence at finite $\Mpl$ has been more limited. In this paper we demonstrated that for general speed of sound models, this expansion is controlled by $\epsilon \ll 1$. In the process, we derived the first and second order corrections to all orders in $\pi$.

One unexpected consequence of the scaling behavior of the gravitational corrections is that they only remain small when compared to the naive dimensional analysis expectations for the higher dimension operators. Specifically, if we assume that all interactions are controlled by a single scale $\Lambda = f_\pi c_s$, then it follows that the gravitational terms are always slow-roll suppressed. If, the higher order interactions were more weakly coupled, the gravitational corrections could ultimately dominate. The importance of this scaling has been previously been seen in the context of sum rules~\cite{Baumann:2015nta,Baumann:2019ghk} and soft theorems~\cite{Green:2022slj} for the inflaton.

Dynamical gravity remains an essential component in how we understand inflationary models. For one, it defines the minimum amount of non-Gaussianity we expect from inflation and sets an aspirational target for future observations~\cite{Cabass:2016cgp}. More subtly, it controls the appearance of non-local terms in the evolution of the scalar fluctuations which would not be permitted in local field theory. These non-local terms also stand in the way of using many conventional results from QFT in the context of inflation. We hope the tools developed in this paper are a step towards a deeper understanding of these gravitational effects and the structure of inflationary predictions beyond the decoupling limit.

\paragraph{Acknowledgements}
We are grateful to Daniel Baumann, Tim Cohen, John McGreevy, Enrico Pajer, and Chia-Hsien Shen for helpful discussions. We are supported by the US~Department of Energy under grant~\mbox{DE-SC0009919}.

\appendix

\section*{Appendices}

\section{Explicit Calculation of the ADM Constraints}\label{app:ADM}

In this appendix, we will review the direct calculation of the leading gravitational correction. The purpose of this discussion is to provide a summary of the conventional approach to these calculation for later comparison with our all-orders result.

We will start from the combined gravititional and inflaton action at zeroth order in the Goldstone boson fluctuations, ${\cal O}(\pi^0)$. At this order, we expect that the constraints are satisfied by $N=1$ and $N^i = 0$, as we would expect from the FRW solution.  To see this explicitly, at linear order in the fluctuations of $N \to N +\delta N$ and $N_i\to N_i +\delta N_i$ (and zeroth order in $\pi$)
\bea
{\cal L}_{\rm EH}+{\cal L}_{\pi,0} &=& a^3 \delta N ( \frac{1}{N^2}( 3 \Mp^2 H^2) - 3 \Mp^2 H^2 - \Mp^2 \dot H + \frac{1}{N^2} \Mp^2 \dot H \nonumber \\ \qquad && - \frac{\Mpl^2}{2 N^2} [  +4 H \partial_i N^i + \frac{1}{2}( \partial_i N_j \partial^i N^j +  \partial_i N_j \partial^j N^i )  - (\partial_i N^i )^2] )  \\ && \qquad + \frac{\Mpl^2}{2} \delta N_i (\partial^2 N^i - \partial^i \partial_j N^j)\nonumber \ .
\eea
Satisfying the constrain equations (equations of motion) requires that the linear terms in the fluctuations vanish. For completeness (and to understand later results), we would like to show that FRW is indeed the correct solution.  Here we will split $N^i = \partial^i \chi + \Psi^i$ where $\partial_i \Psi^i = 0$.  Let's start with the $\delta N^i$ term
\bea
\partial^2 N^i - \partial^i \partial_j N^j = 0 \to \partial^2 \Psi^i = 0 \ .
\eea
The $\delta N$ equations of motion then yield
\beq
 - 2 \Mp^2 H \partial^2 \chi + \left(\frac{1}{N^2} -1 \right) \left(3 \Mp^2 H^2+ \Mp^2 \dot H \right) = 0 \ .
\eeq
Clearly one solution to this equation is $N=1$ and $\chi=0$, which is again just a restatement that FRW is a consistent background solution. This is not a unique solution to these equations, as we can generate other solutions by a change of coordinates.

\vskip 8pt
\noindent {\it Perturbations:} \hskip 6pt  Now we would like to consider the solutions to the constraint equations for $\delta N^i$ and $\delta N$ at first order in perturbation theory. We will expand the Lagrangian ${\cal L} = {\cal L}_{\rm EH}+{\cal L}_{\pi}$ to quadratic order in fluctuations of the metric, ${\cal L}_{\delta g, 2}$, dropping terms that do not contain $N$ or $N^i$, 
\bea
{\cal L}_{\delta g, 2} &=&  \Mp^2 (- 3 H^2 \delta N^2 -2 H \delta N \partial_i N^i ) \nonumber \\
&&- \Mp^2 \dot H(-2 \delta N \dot \pi -2 N^i \partial_i \pi -  (\delta N)^2)+ 2 M_2^4( \delta N^2 - 2 \delta N \dot \pi) \\
&&+\frac{\Mp^2}{2} [\frac{1}{2}( \partial_i N_j \partial^i N^j +  \partial_i N_j \partial^j N^i )  - (\partial_i N^i )^2 ]-\delta N \Mp^2 ( 6 H \dot H \pi + \ddot H \pi) \ . \nonumber
\eea
Again, we will find the equations of motion for $\delta N$ and $N^i$ that this order and use them to determine the corrections to the metric.  The equations of motion for $N^i$ are
\bea
&&2 \Mp^2 H \partial_i \delta N + 2 \Mp^2 \dot H \partial_i \pi - \frac{\Mp^2}{2} (\partial^2 \Psi_i) = 0  \\
&& \to 2 \Mp^2 H \partial_i (\delta N - H \epsilon \pi) - \frac{\Mp^2}{2} (\partial^2 \Psi_i) = 0 \ .
\eea
Working in momentum space, let us assume that $\delta N - H \epsilon \pi = f(\k)$ for some unknown function $f(\k)$.  Since $k^i \Psi_i = 0$, any solution for non-zero $\Psi$ would require $k^i f(k) k^{-2} = 0$.  Therefore, we have $\delta N = H\epsilon \pi $ and $\Psi_i = 0$ for any finite $\k \neq 0$.

Next, the equations of motion for $\delta N$ give
\begin{align}
- 6 \Mp^2 H^2 \delta N - 2 \Mp^2 H \partial^2 \chi + 2 \Mp^2 \dot H \dot \pi + 2 \Mp^2 \dot H \delta N & \nonumber \\
+ 4 M_2^4 \delta N - 4 M_2^4 \dot \pi- 6 \Mp^2 H \dot H \pi &= 0 \nonumber \\ 
\to  - 2 \Mp^2 H \partial^2 \chi + 2 \Mp^2 \dot H \dot \pi + 2 \Mp^2 \dot H H \epsilon \pi + 4 M_2^4 H \epsilon \pi - 4 M_2^4 \dot \pi &= 0
\end{align}
The solution to these equations at order $\epsilon$ (i.e. dropping $\epsilon^2$ terms) is
\beq
\partial^2 \chi = - \frac{H \epsilon \dot \pi}{c_s^2} \ ,
\eeq
where we used $2 M_2^4 - \Mp^2 \dot H = - \Mp^2 \dot H c_s^{-2}$, or equivalently $M_2^4 =- \tfrac{1}{2}\Mp^2 \dot H c_s^{-2}(1- c_s^2)$.

Now, let's plug this solution back in the action, again only keeping terms to order $\epsilon$.  Using our solutions, 
\beq
\delta N = \epsilon H \pi \qquad \partial^i N_i = - \frac{H \epsilon \dot \pi}{c_s^2} \ ,
\eeq
we find the correction to the quadratic action becomes
\bea
{\cal L}_{\delta g, 2}&\to&  \Mp^2 (- 3 H^2 \delta N^2 -2 H \delta N \partial_i N^i ) - \Mp^2 \dot H(-2 \delta N \dot \pi + 2 \partial_i N^i  \pi )\nonumber \\
&&- 2 M_2^4(2 \delta N \dot \pi) - 6 \Mp^2  H \dot H\delta N \pi  - 3 \Mp^2 \dot H^2 \pi^2  \\
&=&  \Mp^2 (+ 3 \dot H \epsilon H^2 \pi^2 - 2 \dot H \pi H \frac{\epsilon \dot \pi}{c_s^2} )  + \frac{\Mp^2 \dot H}{c_s^2}(2 H\epsilon \pi \dot \pi) \\
&&+\Mp^2 \dot H 2H \frac{\epsilon\dot \pi}{c_s^2} \pi - 3 \Mp^2\dot H  H^2 \epsilon \pi^2 \nonumber \\
&=& - 3 \frac{\Mp^2 \dot H}{c_s^2} H^2 \epsilon \pi^2
\eea
The last term in the first equation comes from expanding $3\Mp^2 H^2$ to second order in $\pi$. In the last line, we dropped a total derivatives. The important this is that when we add this contribution back into the quadratic action for $\pi$, we have
\beq
{\cal L}_{2} = {\cal L}_{\delta g, 2} + {\cal L}_{\pi,2} \to -  \frac{\Mp^2 \dot H}{c_s^2} (\dot \pi^2 - c_s^2 \frac{\partial_i \pi \partial^i \pi}{a^2} + 3 H^2 \epsilon \pi^2 ) \ .
\eeq
At horizon crossing, $\dot \pi^2  = H^2 \pi^2 \gg 3 \epsilon H^2 \pi^2$ so the mixing with gravity is a small contribution (it is slow roll suppressed).

\section{Stress Tensor of the Inflaton}\label{app:stress}

The stress tensor plays a major role in the results of this paper. In this appendix, we derived some of the key results regarding the stress tensor for the Goldstone boson in the EFT of inflation.

Give an action $S_{\rm matter}$ for any matter (and/or radiation) fields, one can define a stress tensor via
\beq
T_{\mu \nu}=\frac{-2}{\sqrt{-g}} \frac{\delta S_{\text {matter }}}{\delta g^{\mu \nu}}=\frac{-2}{\sqrt{-g}} \frac{\partial\left(\sqrt{-g} \mathcal{L}_{\text {matter }}\right)}{\partial g^{\mu \nu}}=-2 \frac{\partial \mathcal{L}_{\text {matter }}}{\partial g^{\mu \nu}}+g_{\mu \nu} \mathcal{L}_{\text {matter }} \ .
\eeq
When coupled to gravity, so that $S = S_{\rm EH}+ S_{\rm matter}$ where $S_{\rm EH}$ is the Einstein-Hilbert action, this definition gives rise to Einstein's field equations for the metric.

In single-field inflation, the action for the inflaton can be described in terms of the Goldstone boson, $\pi$, such that role so $S_{\rm matter}$ is played by
\begin{align}
S_\pi &=\int d^4 x \sqrt{-g} \sum_{n=0}^{\infty}  \frac{M^4_n(U)}{n!}  (\partial_\mu U \partial^\mu U+1)^n + \ldots \\
&=\int d^4 x \sqrt{-g} \sum_{n=0}^{\infty} \frac{M^4_n(t+\pi)}{n!}  (-2\dot \pi + \partial_\mu \pi \partial^\mu \pi)^n + \ldots \ .
\end{align}
As a result,  the stress-tensor of the inflaton is given by
\beq
T_{\mu \nu} = - 2 \partial_\mu U \partial_\nu U \sum_{n=0}^{\infty} \frac{M^4_{n+1}(U)}{n!} (\partial_\mu U \partial^\mu U+1)^n  + g_{\mu \nu} \sum_{n=0}^{\infty}  \frac{M^4_n(U)}{n!}  (\partial_\mu U \partial^\mu U+1)^n \ .
\eeq
When we evaluate the stress tensor with $\pi=0$, we get the stress tensor responsible for the background FRW metric, $T_{\mu \nu}(\pi=0) = \bar T_{\mu \nu}$. In the EFT of inflation, this is
\beq
\bar T_{00} = -2 M_1^4 - M_0^4 = \rho \qquad \bar T_{ii} =a^2 M_0^4 = a^2 p
\eeq
We can now use these in the Friedmann equations to determine $H(t)$ in terms of $M_0^4(t)$ and $M_1^4(t)$, or vice versa, via
\beq
\begin{aligned}
& 3 \Mp^2 H^2=3 \Mp^2\left(\frac{\dot{a}}{a}\right)^2= \rho = -2 M_1^4 - M_0^4 \\
&6 \Mp^2( \dot{H}+H^2) =6 \Mp^2 \left( \frac{\ddot{a}}{a} \right)=-\rho-3 p = M_0^4 +2 M_1^4 - 3 M_0^4 \ .
\end{aligned}
\eeq
These equations are solved to give 
\beq
M_1^4= \Mp^2 \dot H \qquad M_0^4 = - \Mp^2(3H^2 + 2 \dot H)  \  .
\eeq
This is the sense in which $M_0^4$ and $M_1^4$ are universal quantities in the EFT of inflation.

We can defined the fluctuations of the stress tensor via $T_{\mu \nu} = \bar T_{\mu \nu} + \delta T_{\mu \nu}$. At linear order in $\pi$, $\delta T_{\mu \nu}$ is given by
\bea
\delta T_{\mu \nu} &=& - 2 M_1^4(t)  (\delta_{0 \mu} \partial_\nu \pi + \delta_{0 \nu}  \partial_\mu \pi)-2 \delta_{\mu 0} \delta_{\nu 0}\left( \dot M_1^4(t) \pi + M_2^4(t) (-2 \dot \pi)  \right)\\
&&+ g_{\mu \nu}\left( \dot M_0^4(t) \pi - 2 \dot \pi M_1^4(t) \right)  \ .
\eea
Using $M_0^4$ and $M_1^4$ in terms of the background, and $M_2^4$ in terms of the speed of sound,
\beq
M_2^4 =- \tfrac{1}{2}\Mp^2 \dot H c_s^{-2}(1- c_s^2) \ ,
\eeq
we find the linear contributions to the stress tensor are
\bea
\delta T_{00} 
&=& - 2 \frac{\Mp^2 \dot H}{c_s^{2}} \dot \pi +6 \Mpl^2 H \dot H \pi + {\cal O}(\pi^2) \ ,
\eea
and
\beq
\delta T_{0j} = - 2 \Mp^2 \dot H \partial_j \pi+ {\cal O}(\pi^2) \ .
\eeq
The results play crucial roles in the defining $\pi$ as a Goldstone  boson with decay constant $f_\pi$.

For the purpose of imposing the ADM constraints, it is useful to rewrite these expressions in terms of derivatives with respect to $N$ and $N^i$. From the definition, 
\beq
T_{\mu \nu}=\frac{-2}{\sqrt{-g}} \frac{\delta S_{\pi}}{\delta g^{\mu \nu}}
\eeq
we can apply the chain rule as
\beq
\frac{\delta S_{\pi}}{\delta N} = \frac{\delta S_{\pi}}{\delta g^{\mu \nu}} \frac{ \delta g^{\mu \nu}}{\delta N} \qquad \frac{\delta S_{\pi}}{\delta N^i} = \frac{\delta S_{\pi}}{\delta g^{\mu \nu}} \frac{\delta g^{\mu \nu}}{\delta N^i} \ .
\eeq
Now using
\beq
g^{\mu \nu } = \begin{pmatrix}
-\frac{1}{N^2} & \frac{N^i}{N^2} \\
\frac{N^{j}}{N^2} & \frac{\delta_{ij}}{a^2} - \frac{N^i N^j}{N^2} 
\end{pmatrix} \ , 
\eeq
we find
\beq
\frac{\delta S_{\pi}}{\delta N} = \frac{-2}{N^3} \left( - \frac{\delta S_{\pi}}{\delta g^{0 0}} + \frac{\delta S_{\pi}}{\delta g^{0 i}} N^i + \frac{\delta S_{\pi}}{\delta g^{j 0}} N^j - N^i N^j  \frac{\delta S_{\pi}}{\delta g^{i j} }\right)
\eeq
and
\beq
\frac{\delta S_{\pi}}{\delta N^j} = \frac{1}{N^2} \left( \frac{\delta S_{\pi}}{\delta g^{0 j}} + \frac{\delta S_{\pi}}{\delta g^{j 0}} -N^i \frac{\delta S_{\pi}}{\delta g^{i j} }-N^i \frac{\delta S_{\pi}}{\delta g^{ j i } }\right)
\eeq
Rewriting the definition of the stress tensor as
\beq
\frac{\delta S_{\pi}}{\delta g^{\mu \nu}}  = -\frac{\sqrt{-g}}{2} T_{\mu \nu} \ ,
\eeq
we can write the RHS of these expressions of the stress tensor 
\beq \label{$T_00$}
\frac{\delta S_{\pi}}{\delta N} = \frac{\sqrt{-g}}{N^3} \left( - T_{00} + T_{0i} N^i + T_{j0} N^j - N^i N^j T_{ij} \right)
\eeq
and
\beq
\frac{\delta S_{\pi}}{\delta N^j} =- \frac{\sqrt{-g}}{2 N^2} \left( T_{0j} + T_{0j}-N^i T_{ij}-N^i T_{ji} \right) \ .
\eeq
Our main use of this formula will be for derivative evaluated on the fixed background metric, where $N=1$ and $N_i = 0$
\bea
\frac{\delta S_{\pi}}{\delta N}|_{N=1,N^i=0} &=& - a^3 T_{00} \\
\frac{\delta S_{\pi}}{\delta N^j}|_{N=1,N^i=0} &=&- \frac{a^3}{2 } \left( T_{0j} + T_{0j} \right)
\eea
We can then derive the constraint equations, via functional derivatives
\beq
\frac{\delta S_{\text{EH}}}{\delta N}= -\frac{\delta S_{\pi}}{\delta N} \ .
\eeq
We use these results to all orders in $\pi$ using the stress tensor, rather than by brute force expansion of the action.

\section{Mixing with Gravity at Second Order}\label{app:esp2}

\subsection*{Setting up the equations}
For the matter action, we have the following :
\begin{equation}
	\begin{aligned}
	\frac{\delta S_\pi}{\delta N} &= -N^{-2} a^3 (T_{00} - 2 T_{0i} N^i + N^i N^j T_{ij})\\
	\frac{\delta S_\pi}{\delta N^j} &= -\frac{a^3}{N} (T_{0j} - N^i T_{ij})
	\end{aligned}
\end{equation}
The Einstein-Hilbert action, on the other hand, gives :
\begin{equation}
	\frac{\delta S_G}{\delta N} = -\Mpl^2 a^3\frac{1}{2N^2} (E_{ij} E^{ij} - E^2) \hspace{1cm} \frac{\delta S_G}{\delta N^j} = \Mpl^2 a^3 \nabla_i \bigg[E^i_{\,\,j} - \delta^i_{\,\,j} E \bigg] \ .
\end{equation}
Before solving for `second order', we need to define what is the order that we are counting in. We define 'first order' by quantities which depend on ratio of $M_1^{4}, M_2^{4}, \hdots$ with $\Mpl^2 H^2$. For our case, this means that both $\epsilon$ and $\epsilon/c_s^2$ are first order. With this, we can expand each quantity from the variation of the matter action to second order. Furthermore, since $T_{\mu \nu}$ only contains explicit linear dependence on $M_n^4$, all higher orders terms in our power counting arise from the dependence of the stress tensor on $\delta N$ and $N^i$. It is therefore useful to write the stress tensor as an expansion in powers of the lapse and shift that correspond to a given order in $\epsilon$:
\beq
T_{\mu \nu} = \sum_{n=0}^\infty T^{(n)}_{\mu \nu} \qquad T^{(n)}_{\mu \nu} = {\cal O}\left( \epsilon^n \right) \qquad M_n^4,\delta N, N^i ={\cal O}(\epsilon) \qquad \pi = {\cal O}(1) \ .
\eeq
Using this definition, we can expand the terms in the constraint equations to the desired order in the lapse and shift,
\begin{equation}
	\begin{aligned}
    N^{-2} a^3 &(T_{00} - 2T_{0i}N^i + N^i N^j T_{ij})\\
	&\to a^3(3 \overline{T}^{(0)}_{00} (\delta N)^2 - 2 \delta N \delta T_{00}^{(1)} + \delta T_{00}^{(2)} - 2 N^i T_{0i}^{(1)} + N^i N^j \overline{T}^{(0)}_{ij} )\\[6pt]
	-N^{-1} a^3 &(T_{0j} - N^i T_{ij})
	\to a^3( -T_{0j}^{(2)} + \delta N T_{0j}^{(1)} + N^i T_{ij}^{(1)} - \delta N N^i \overline{T}^{(0)}_{ij}) \ .
	\end{aligned}
\end{equation}
Technically, $\delta N = \sum_1^\infty \delta N^{(n)}$ where $\delta N^{(n)} = {\cal O}(\epsilon^n)$, and similarly for $N^i$. At second order in $\epsilon$, the contributions to the stress tensor only require keeping $n=1$ terms in $N$ and $N^i$ and therefore $\delta N \equiv \delta N^{(1)}$ unless otherwise specified.

Next we need to find the Einstein-Hilbert terms to second order. Second order terms here are much easier to define - product of first order terms will be second order, since the gravitational action is made up entirely of metric perturbations. Recall that
\begin{align}
 (E_{ij}E^{ij} - E^2) 
&=  -6 H^2 +4 H \partial_i N^i + \frac{1}{2}( \partial_i N_j \partial^i N^j +  \partial_i N_j \partial^j N^i )  - (\partial_i N^i )^2  
\end{align}
such that expanding to second-order gives the contribution
\begin{align}
	  - &\left(\frac{1}{2N^2} (E_{ij} E^{ij} - E^2) \right) 
   \to -6 H^2 \delta N^{(2)} - 2 H \partial_i N^{i,(2)}+ 4H \delta N \partial_i N^i  \\
   &\qquad\qquad + 9 H^2 (\delta N)^2
	   - \frac{1}{4} (\partial^i N_j \partial_i N^j + \partial^i N_j \partial_j N^i - 2 \partial_i N^i \partial_j N^j) \ .
\end{align}
Varying the action with respect to $\delta N$ and $N^i$ yields the constraint equations
\begin{align}
	\Mpl^2  \bigg[- 6 H^2 \delta N^{(2)} &-2 H \partial_i N^{i,(2)} + 4 H \delta N \partial_i N^i+9 H^2 (\delta N)^2  \nonumber\\
	   &-\frac{1}{4} (\partial^i N_j \partial_i N^j + \partial^i N_j \partial_j N^i - 2 \partial_i N^i \partial_j N^j) \bigg]  \\
	   &= 3 \overline{T}^{(0)}_{00} (\delta N)^2 - 2 \delta N \delta T_{00}^{(1)} + \delta T_{00}^{(2)} - 2 N^i T_{0i}^{(1)} + N^i N^j \overline{T}^{(0)}_{ij} \nonumber \\
	\Mpl^2  \bigg[ - 2H \partial_j \delta N^{(2)} &+ \frac{1}{2}(\partial^2 N_j^{(2)} - \partial^j \partial_i N^{i,(2)}) - \frac{1}{2}(\delta N) ( \partial^2 N_j - \partial^j \partial_i N^i) \nonumber \\
	& - 2 H \delta N \, \partial_j (\delta N) - \frac{1}{2}\partial_i (\delta N) (\partial^i N_j + \partial^j N_i - 2 \delta_{ij} \partial^k N_k) \bigg] \\
	 & = T_{0j}^{(2)} - \delta N T_{0j}^{(1)} - N^i T_{ij}^{(1)} + \delta N N^i \overline{T}^{(0)}_{ij} \ . \nonumber	 
\end{align}
This equation yields the second order constraint to all orders in $\pi$. However, in the interest of finding explicit solutions, we will use it to extract the ${\cal O}(\pi^2)$ solution.  
\subsection*{Solving for the Lapse}

First we solve for $\delta N$. We do this by solving the second equation, which reads :
\begin{equation}
	\begin{aligned}
	- 2 \Mpl^2 H \partial_j \delta N^{(2)} &+ \frac{\Mpl^2}{2}(\partial^2 N_j^{(2)} - \partial^j \partial_i N^{i,(2)}) - \frac{\Mpl^2}{2}(\delta N) (\partial^2 N_j - \partial^j \partial_i N^i) \\
	& - 2 \Mpl^2 H \delta N \, \partial_j (\delta N) - \frac{\Mpl^2}{2}\partial_i (\delta N) (\partial^i N_j + \partial^j N_i - 2 \Mpl^2\delta_{ij} \partial^k N_k)  \\
	 & = T_{0j}^{(2)} - \delta N T_{0j}^{(1)} - N^i T_{ij}^{(1)} + \delta N N^i \overline{T}^{(0)}_{ij}
	\end{aligned}
\end{equation}
Now, we have to find the right hand side to second order in $\pi$. We have:
\begin{equation}
	\begin{aligned}
	\delta T_{0j}^{(2)}  =& -2 (1 + \dot{\pi}) \partial_j \pi M_2^{4}(2 \delta N ) + a^2 N_j \left ( M_0^{4} + M_1^{4} (-2\dot{\pi}) \right )\\
	=& 2 \delta N \frac{\Mpl^2 \dot{H}}{c_s^2}(1 - c_s^2) \partial_j \pi - 6 a^2 \Mpl^2 H \dot{H} \pi N_j - 2 a^2 \Mpl^2 \dot{H} N_j - 2 a^2 N_j \Mpl^2 \dot{H} \dot{\pi}\\
	T_{0j}^{(1)} &= -2 \Mpl^2 \dot{H} \partial_j \pi - 3 a^2 N_j \Mpl^2 H^2\\
	\delta T_{ij}^{(1)} =& -2 \Mpl^2 \dot{H} \partial_i \pi \partial_j \pi + a^2\delta_{ij} (-3\Mpl^2 H^2 - 6\Mpl^2 H \dot{H} \pi - 2\Mpl^2\dot{H}) \\&
 + \Mpl^2 \dot{H} a^2 \delta_{ij} ( - 2\dot{\pi}) 
	\end{aligned}
\end{equation}
Then, we have that
\begin{equation}
\begin{aligned}
	\delta T_{0j}^{(2)} - \delta N T_{0j}^{(1)} - N^i T_{ij}^{(1)} + \delta N N^i \overline{T}^{(0)}_{ij} &=  2 \frac{\Mpl^2 \dot{H}}{c_s^2}(1 - c_s^2) \delta N \partial_j \pi + 2 \Mpl^2 \dot{H} \delta N  \partial_j \pi \\
	&= 2\Mpl^2 \dot{H} \frac{\epsilon}{c_s^2} H \pi \partial_j \pi
\end{aligned}
\end{equation}
Now, dividing the original equation by $\Mpl^2 H^2$, we get :
\begin{equation}
	\begin{aligned}
	-2 H^{-1}\partial^2 \delta N^{(2)} &- 2 H^{-1} \partial_j (\delta N \partial_j (\delta N)) - \frac{H^{-2}}{2} \partial_j \bigg[\partial_i (\delta N) (\partial_i \partial_j \chi - \delta_{ij}\partial^2 \chi) \bigg] \\
	&= -2 \frac{\epsilon^2}{c_s^2} H \partial_j (\pi \partial_j \pi)\\
	-2H^{-1} \partial^2 \delta N^{(2)} &- 2 \epsilon^2 H \partial_j (\pi \partial_j \pi) + \frac{\epsilon^2}{2c_s^2} \partial_j \bigg[\partial_i \pi \left ( \frac{\partial_i \partial_j}{\partial^2} \dot{\pi} - \delta_{ij}\dot{\pi} \right ) \bigg]\\
	&= -2\frac{\epsilon^2}{c_s^2} H \partial_j (\pi \partial_j \pi) 
	\end{aligned}
\end{equation}
Finally, we can solve for $\delta N^{(2)}$. We get :
\begin{equation}
	\begin{aligned}
	\delta N^{(2)} &= \frac{\partial_j}{\partial^2} \bigg [ \frac{\epsilon^2}{c_s^2}(1 - c_s^2) H^2 \pi \partial_j \pi - \frac{\epsilon^2}{c_s^2} \frac{H}{4} \left (\dot{\pi}\partial_j \pi - \partial_i \pi \frac{\partial_i \partial_j}{\partial^2}  \dot{\pi} \right ) \bigg]\\
	\end{aligned}
\end{equation}
This is the solution for $\delta N^{(2)}$. We can see clearly that at $ {\cal{O}} (\pi^2)$, we have that $\delta N^{(2)} \sim \epsilon^2/c_s^2$. Now, let us solve for $\chi^{(2)}$. 
\subsection*{Solution for the Shift}
We now use our $\delta N^{(2)}$ solution to solve for $\partial_i N^i = a^{-2} \partial^2 \chi^{(2)}$. Now, the equation we are after is :
\begin{equation}
	\begin{aligned}
	-6 \Mpl^2 H^2 \delta N^{(2)} &- 2 \Mpl^2 \frac{H}{a^2} \partial_i N^{i,(2)} + 4 \Mpl^2 \frac{H}{a^2} \delta N \partial_i N^i + 9 \Mpl^2 H^2 (\delta N)^2\\
	&- \frac{\Mpl^2}{4a^4} (\partial^i N_j \partial_i N^j + \partial^i N_j \partial_j N^i - 2 \partial_i N^i \partial_j N^j)\\
	&= ( 3 \overline{T}^{(0)}_{00} (\delta N)^2 - 2 \delta N \delta T_{00}^{(1)} + \delta T_{00}^{(2)} - 2 N^i T_{0i}^{(1)} + N^i N^j \overline{T}^{(0)}_{ij}  )\\
	\end{aligned}
\end{equation}
Again, we start with finding the right hand side. The first term is clear :
\begin{equation}
	 3 \overline{T}^{(0)}_{00} (\delta N)^2 = 3\Mpl^2 \epsilon^2 H^2 \pi^2
\end{equation}
We have that : 
\begin{equation}
	\begin{aligned}
	\delta T_{00}^{(2)} =& -2 (1 + 2\dot{\pi}) M_{2}^{4} \left( 2 N^i \partial_i \pi + 2 \delta N (1 + 2 \dot{\pi}) \right)\\
	&- \bigg[ M_1^{4} (2 \delta N (1 + 2\dot{\pi}) + 2N^i \partial_i \pi) + M_2^{4} (2 \delta N - 2 \dot{\pi})^2 \bigg] \\
	&+ \bigg[ - 2\delta N - (\delta N)^2 + a^2 N^i N_i \bigg] \left ( M_0^{4} + M_1^{4}(-2\dot{\pi})  \right )\\
    =& 2 \Mpl^2 \dot{H} c_s^{-2} \delta N + 2 \Mpl^2 \dot{H}c_s^{-2} N^i \partial_i \pi - 4\Mpl^2 \dot{H} N^i \partial_i \pi + 12 \Mpl^2 H \dot{H} \pi \delta N\\
    &+ 6 \Mpl^2 \dot{H} c_s^{-2} \dot{\pi} \delta N - 6 \Mpl^2 \dot{H} \dot{\pi} \delta N + 3 \Mpl^2 H^2 (\delta N)^2 - 3 a^2\Mpl^2 H^2 N^i N_i
	\end{aligned}
\end{equation}
We also have : 
\begin{equation}
	\begin{aligned}
	\delta T_{00}^{(1)} =& -2(1 + \dot{\pi})^2 \bigg[\Mpl^2 \dot{H} - \frac{1}{2} \Mpl^2 \dot{H} c_s^{-2}(1 - c_s^2) (-2\dot{\pi}) \bigg] + 6\Mpl^2 H \dot{H} \pi\\
	 & + 2 \Mpl^2 \dot{H} + M_1^{4} (-2 \dot{\pi}) + 2\delta N (3 \Mpl^2 H^2)\\
	=&\,  6 \Mpl^2 H^2 \delta N - 2\frac{\Mpl^2 \dot{H} \dot{\pi}}{c_s^2} + 6 \Mpl^2 H \dot{H} \pi
	\end{aligned}
\end{equation}
Finally, we have :
\begin{equation}
	\delta T_{0j}^{(1)} = -2 \Mpl^2 \dot{H} \partial_j \pi - 3 N_j \Mpl^2 H^2
\end{equation}
Thus, we have that
\begin{equation}
	\begin{aligned}
    &3 \overline{T}^{(0)}_{00} (\delta N)^2 - 2 \delta N \delta T_{00}^{(1)} + \delta T_{00}^{(2)} - 2 N^i T_{0i}^{(1)} + N^i N^j \overline{T}^{(0)}_{ij}\\
	=& 2 \frac{\Mpl^2 \dot{H}}{c_s^2} \delta N + 2\frac{\Mpl^2 \dot{H}}{c_s^2} N^i \partial_i \pi \\
    &+ 10 \Mpl^2 \dot{H} c_s^{-2} \dot{\pi} \delta N - 6 \Mpl^2 \dot{H} \delta N \dot{\pi} - 12 \Mpl^2 H^2 (\delta N)^2 
	\end{aligned}
\end{equation}
Putting all of this together, we see that we have :
\begin{equation}
	\begin{aligned}
	-6 \Mpl^2 H^2 \delta N^{(2)} &- 2 \Mpl^2 \frac{H}{a^2} \partial_i N^{i,(2)} + 4 \Mpl^2 \frac{H}{a^2} \delta N \partial_i N^i + 9 \Mpl^2 H^2 (\delta N)^2\\
	&- \frac{\Mpl^2}{4a^4} (\partial^i N_j \partial_i N^j + \partial^i N_j \partial_j N^i - 2 \partial_i N^i \partial_j N^j)\\
	   &= 3 \overline{T}^{(0)}_{00} (\delta N)^2 - 2 \delta N \delta T_{00}^{(1)} + \delta T_{00}^{(2)} - 2 N^i T_{0i}^{(1)} + N^i N^j \overline{T}^{(0)}_{ij} \\
	   \partial^2 \chi^{(2)} &= -3 H \delta N^{(2)} + 2 \delta N \partial^2 \chi + \frac{9}{2} H (\delta N)^{2} - \frac{1}{4H}((\partial_i \partial_j \chi)^2 - (\partial^2 \chi)^2) \\
	   & + \frac{\epsilon}{c_s^2} H \delta N - \frac{\epsilon}{c_s^2} H N^i \partial_i \pi + 5 \frac{\epsilon}{c_s^{2}} H \dot{\pi} \delta N - 3 \epsilon H \delta N \dot{\pi} + 6 H (\delta N)^2 \\
	 \partial^2 \chi^{(2)} &= -3 H \delta N^{(2)} + \frac{\epsilon^2}{c_s^2}H^2 \pi - \frac{\epsilon^2 H^3}{4 c_s^4}\left ( \left ( \frac{\partial_i \partial_j}{\partial^2} \dot{\pi} \right )^2 - \dot{\pi}^2\right ) + 3 \frac{\epsilon^2}{c_s^2} H^2 \dot{\pi} \pi\\
	 & + \frac{\epsilon^2}{c_s^4} H^2 \frac{\partial_i}{\partial^2} \dot{\pi} \partial^i \pi  - 3\epsilon^2 H^2 \dot{\pi} \pi  + \frac{21}{2} \epsilon^2 H^3 \pi^2
	\end{aligned}
\end{equation}

To summarize, at order ${\cal {O}} (\pi^2)$, we explicitly get :

\begin{equation}
	\begin{aligned}
	\delta N^{(2)} &= \frac{\partial_j}{\partial^2} \bigg [ \frac{\epsilon^2}{c_s^2}(1 - c_s^2) H^2 \pi \partial_j \pi - \frac{\epsilon^2}{c_s^2} \frac{H}{4} \left (\dot{\pi}\partial_j \pi - \partial_i \pi \frac{\partial_i \partial_j}{\partial^2}  \dot{\pi} \right ) \bigg]\\
	\partial^2 \chi^{(2)} &= -3 H \delta N^{(2)} + \frac{\epsilon^2}{c_s^2}H^2 \pi - \frac{\epsilon^2 H^3}{4 c_s^4}\left ( \left ( \frac{\partial_i \partial_j}{\partial^2} \dot{\pi} \right )^2 - \dot{\pi}^2\right ) + 3 \frac{\epsilon^2}{c_s^2} H^2 \dot{\pi} \pi\\
	 & + \frac{\epsilon^2}{c_s^4} H^2 \frac{\partial_i}{\partial^2} \dot{\pi} \partial^i \pi  - 3\epsilon^2 H^2 \dot{\pi} \pi  + \frac{21}{2} \epsilon^2 H^3 \pi^2
	\end{aligned}
\end{equation}

\phantomsection
\addcontentsline{toc}{section}{References}
\small
\bibliographystyle{utphys}
\bibliography{GBRefs}

\providecommand{\href}[2]{#2}\begingroup\raggedright\begin{thebibliography}{10}

\bibitem{Achucarro:2022qrl}
A.~Ach\'ucarro {\em et~al.}, ``{Inflation: Theory and Observations},''
  \href{http://arxiv.org/abs/2203.08128}{{\ttfamily arXiv:2203.08128
  [astro-ph.CO]}}.

\bibitem{Planck:2019kim}
{\bfseries Planck} Collaboration, Y.~Akrami {\em et~al.}, ``{Planck 2018
  results. IX. Constraints on primordial non-Gaussianity},''
  \href{http://dx.doi.org/10.1051/0004-6361/201935891}{{\em Astron. Astrophys.}
  {\bfseries 641} (2020) A9}, \href{http://arxiv.org/abs/1905.05697}{{\ttfamily
  arXiv:1905.05697 [astro-ph.CO]}}.

\bibitem{SPHEREx:2014bgr}
{\bfseries SPHEREx} Collaboration, O.~Dor\'e {\em et~al.}, ``{Cosmology with
  the SPHEREX All-Sky Spectral Survey},''
  \href{http://arxiv.org/abs/1412.4872}{{\ttfamily arXiv:1412.4872
  [astro-ph.CO]}}.

\bibitem{DESI:2022lza}
{\bfseries DESI} Collaboration, D.~J. Schlegel {\em et~al.}, ``{A Spectroscopic
  Road Map for Cosmic Frontier: DESI, DESI-II, Stage-5},''
  \href{http://arxiv.org/abs/2209.03585}{{\ttfamily arXiv:2209.03585
  [astro-ph.CO]}}.

\bibitem{Schlegel:2022vrv}
D.~J. Schlegel {\em et~al.}, ``{The MegaMapper: A Stage-5 Spectroscopic
  Instrument Concept for the Study of Inflation and Dark Energy},''
  \href{http://arxiv.org/abs/2209.04322}{{\ttfamily arXiv:2209.04322
  [astro-ph.IM]}}.

\bibitem{Chang:2022lrw}
C.~L. Chang {\em et~al.}, ``{Report of the Topical Group on Cosmic Frontier 5
  Dark Energy and Cosmic Acceleration: Cosmic Dawn and Before for Snowmass
  2021},'' in {\em {Snowmass 2021}}.
\newblock 9, 2022.
\newblock \href{http://arxiv.org/abs/2209.08265}{{\ttfamily arXiv:2209.08265
  [hep-ex]}}.

\bibitem{Maldacena:2002vr}
J.~M. Maldacena, ``{Non-Gaussian features of primordial fluctuations in single
  field inflationary models},''
  \href{http://dx.doi.org/10.1088/1126-6708/2003/05/013}{{\em JHEP} {\bfseries
  05} (2003) 013}, \href{http://arxiv.org/abs/astro-ph/0210603}{{\ttfamily
  arXiv:astro-ph/0210603}}.

\bibitem{Creminelli:2004yq}
P.~Creminelli and M.~Zaldarriaga, ``{Single field consistency relation for the
  3-point function},''
  \href{http://dx.doi.org/10.1088/1475-7516/2004/10/006}{{\em JCAP} {\bfseries
  10} (2004) 006}, \href{http://arxiv.org/abs/astro-ph/0407059}{{\ttfamily
  arXiv:astro-ph/0407059}}.

\bibitem{Bardeen:1980kt}
J.~M. Bardeen, ``{Gauge Invariant Cosmological Perturbations},''
  \href{http://dx.doi.org/10.1103/PhysRevD.22.1882}{{\em Phys. Rev. D}
  {\bfseries 22} (1980) 1882--1905}.

\bibitem{Salopek:1990jq}
D.~S. Salopek and J.~R. Bond, ``{Nonlinear evolution of long wavelength metric
  fluctuations in inflationary models},''
  \href{http://dx.doi.org/10.1103/PhysRevD.42.3936}{{\em Phys. Rev. D}
  {\bfseries 42} (1990) 3936--3962}.

\bibitem{Lyth:2002my}
D.~H. Lyth, C.~Ungarelli, and D.~Wands, ``{The Primordial density perturbation
  in the curvaton scenario},''
  \href{http://dx.doi.org/10.1103/PhysRevD.67.023503}{{\em Phys. Rev. D}
  {\bfseries 67} (2003) 023503},
  \href{http://arxiv.org/abs/astro-ph/0208055}{{\ttfamily
  arXiv:astro-ph/0208055}}.

\bibitem{Zaldarriaga:2003my}
M.~Zaldarriaga, ``{Non-Gaussianities in models with a varying inflaton decay
  rate},'' \href{http://dx.doi.org/10.1103/PhysRevD.69.043508}{{\em Phys. Rev.
  D} {\bfseries 69} (2004) 043508},
  \href{http://arxiv.org/abs/astro-ph/0306006}{{\ttfamily
  arXiv:astro-ph/0306006}}.

\bibitem{Sasaki:2006kq}
M.~Sasaki, J.~Valiviita, and D.~Wands, ``{Non-Gaussianity of the primordial
  perturbation in the curvaton model},''
  \href{http://dx.doi.org/10.1103/PhysRevD.74.103003}{{\em Phys. Rev. D}
  {\bfseries 74} (2006) 103003},
  \href{http://arxiv.org/abs/astro-ph/0607627}{{\ttfamily
  arXiv:astro-ph/0607627}}.

\bibitem{Chen:2009zp}
X.~Chen and Y.~Wang, ``{Quasi-Single Field Inflation and Non-Gaussianities},''
  \href{http://dx.doi.org/10.1088/1475-7516/2010/04/027}{{\em JCAP} {\bfseries
  04} (2010) 027}, \href{http://arxiv.org/abs/0911.3380}{{\ttfamily
  arXiv:0911.3380 [hep-th]}}.

\bibitem{Senatore:2010wk}
L.~Senatore and M.~Zaldarriaga, ``{The Effective Field Theory of Multifield
  Inflation},'' \href{http://dx.doi.org/10.1007/JHEP04(2012)024}{{\em JHEP}
  {\bfseries 04} (2012) 024}, \href{http://arxiv.org/abs/1009.2093}{{\ttfamily
  arXiv:1009.2093 [hep-th]}}.

\bibitem{Baumann:2011nk}
D.~Baumann and D.~Green, ``{Signatures of Supersymmetry from the Early
  Universe},'' \href{http://dx.doi.org/10.1103/PhysRevD.85.103520}{{\em Phys.
  Rev. D} {\bfseries 85} (2012) 103520},
  \href{http://arxiv.org/abs/1109.0292}{{\ttfamily arXiv:1109.0292 [hep-th]}}.

\bibitem{Arkani-Hamed:2015bza}
N.~Arkani-Hamed and J.~Maldacena, ``{Cosmological Collider Physics},''
  \href{http://arxiv.org/abs/1503.08043}{{\ttfamily arXiv:1503.08043
  [hep-th]}}.

\bibitem{Arkani-Hamed:2003juy}
N.~Arkani-Hamed, P.~Creminelli, S.~Mukohyama, and M.~Zaldarriaga, ``{Ghost
  inflation},'' \href{http://dx.doi.org/10.1088/1475-7516/2004/04/001}{{\em
  JCAP} {\bfseries 04} (2004) 001},
  \href{http://arxiv.org/abs/hep-th/0312100}{{\ttfamily arXiv:hep-th/0312100}}.

\bibitem{Alishahiha:2004eh}
M.~Alishahiha, E.~Silverstein, and D.~Tong, ``{DBI in the sky},''
  \href{http://dx.doi.org/10.1103/PhysRevD.70.123505}{{\em Phys. Rev. D}
  {\bfseries 70} (2004) 123505},
  \href{http://arxiv.org/abs/hep-th/0404084}{{\ttfamily arXiv:hep-th/0404084}}.

\bibitem{Chen:2006nt}
X.~Chen, M.-x. Huang, S.~Kachru, and G.~Shiu, ``{Observational signatures and
  non-Gaussianities of general single field inflation},''
  \href{http://dx.doi.org/10.1088/1475-7516/2007/01/002}{{\em JCAP} {\bfseries
  01} (2007) 002}, \href{http://arxiv.org/abs/hep-th/0605045}{{\ttfamily
  arXiv:hep-th/0605045}}.

\bibitem{Creminelli:2006xe}
P.~Creminelli, M.~A. Luty, A.~Nicolis, and L.~Senatore, ``{Starting the
  Universe: Stable Violation of the Null Energy Condition and Non-standard
  Cosmologies},'' \href{http://dx.doi.org/10.1088/1126-6708/2006/12/080}{{\em
  JHEP} {\bfseries 12} (2006) 080},
  \href{http://arxiv.org/abs/hep-th/0606090}{{\ttfamily arXiv:hep-th/0606090}}.

\bibitem{Cheung:2007st}
C.~Cheung, P.~Creminelli, A.~L. Fitzpatrick, J.~Kaplan, and L.~Senatore, ``{The
  Effective Field Theory of Inflation},''
  \href{http://dx.doi.org/10.1088/1126-6708/2008/03/014}{{\em JHEP} {\bfseries
  03} (2008) 014}, \href{http://arxiv.org/abs/0709.0293}{{\ttfamily
  arXiv:0709.0293 [hep-th]}}.

\bibitem{Cornwall:1974km}
J.~M. Cornwall, D.~N. Levin, and G.~Tiktopoulos, ``{Derivation of Gauge
  Invariance from High-Energy Unitarity Bounds on the s Matrix},''
  \href{http://dx.doi.org/10.1103/PhysRevD.10.1145}{{\em Phys. Rev. D}
  {\bfseries 10} (1974) 1145}. [Erratum: Phys.Rev.D 11, 972 (1975)].

\bibitem{Lee:1977eg}
B.~W. Lee, C.~Quigg, and H.~B. Thacker, ``{Weak Interactions at Very
  High-Energies: The Role of the Higgs Boson Mass},''
  \href{http://dx.doi.org/10.1103/PhysRevD.16.1519}{{\em Phys. Rev. D}
  {\bfseries 16} (1977) 1519}.

\bibitem{Chanowitz:1985hj}
M.~S. Chanowitz and M.~K. Gaillard, ``{The TeV Physics of Strongly Interacting
  W's and Z's},'' \href{http://dx.doi.org/10.1016/0550-3213(85)90580-2}{{\em
  Nucl. Phys. B} {\bfseries 261} (1985) 379--431}.

\bibitem{Gounaris:1986cr}
G.~J. Gounaris, R.~Kogerler, and H.~Neufeld, ``{Relationship Between
  Longitudinally Polarized Vector Bosons and their Unphysical Scalar
  Partners},'' \href{http://dx.doi.org/10.1103/PhysRevD.34.3257}{{\em Phys.
  Rev. D} {\bfseries 34} (1986) 3257}.

\bibitem{Yao:1988aj}
Y.-P. Yao and C.~P. Yuan, ``{Modification of the Equivalence Theorem Due to
  Loop Corrections},'' \href{http://dx.doi.org/10.1103/PhysRevD.38.2237}{{\em
  Phys. Rev. D} {\bfseries 38} (1988) 2237}.

\bibitem{Bagger:1989fc}
J.~Bagger and C.~Schmidt, ``{Equivalence Theorem Redux},''
  \href{http://dx.doi.org/10.1103/PhysRevD.41.264}{{\em Phys. Rev. D}
  {\bfseries 41} (1990) 264}.

\bibitem{Veltman:1989ud}
H.~G.~J. Veltman, ``{The Equivalence Theorem},''
  \href{http://dx.doi.org/10.1103/PhysRevD.41.2294}{{\em Phys. Rev. D}
  {\bfseries 41} (1990) 2294}.

\bibitem{Baumann:2011su}
D.~Baumann and D.~Green, ``{Equilateral Non-Gaussianity and New Physics on the
  Horizon},'' \href{http://dx.doi.org/10.1088/1475-7516/2011/09/014}{{\em JCAP}
  {\bfseries 09} (2011) 014}, \href{http://arxiv.org/abs/1102.5343}{{\ttfamily
  arXiv:1102.5343 [hep-th]}}.

\bibitem{Cheung:2007sv}
C.~Cheung, A.~L. Fitzpatrick, J.~Kaplan, and L.~Senatore, ``{On the consistency
  relation of the 3-point function in single field inflation},''
  \href{http://dx.doi.org/10.1088/1475-7516/2008/02/021}{{\em JCAP} {\bfseries
  02} (2008) 021}, \href{http://arxiv.org/abs/0709.0295}{{\ttfamily
  arXiv:0709.0295 [hep-th]}}.

\bibitem{Senatore:2012ya}
L.~Senatore and M.~Zaldarriaga, ``{The constancy of $\zeta$ in single-clock
  Inflation at all loops},''
  \href{http://dx.doi.org/10.1007/JHEP09(2013)148}{{\em JHEP} {\bfseries 09}
  (2013) 148}, \href{http://arxiv.org/abs/1210.6048}{{\ttfamily arXiv:1210.6048
  [hep-th]}}.

\bibitem{Assassi:2012et}
V.~Assassi, D.~Baumann, and D.~Green, ``{Symmetries and Loops in Inflation},''
  \href{http://dx.doi.org/10.1007/JHEP02(2013)151}{{\em JHEP} {\bfseries 02}
  (2013) 151}, \href{http://arxiv.org/abs/1210.7792}{{\ttfamily arXiv:1210.7792
  [hep-th]}}.

\bibitem{Cohen:2020php}
T.~Cohen and D.~Green, ``{Soft de Sitter Effective Theory},''
  \href{http://dx.doi.org/10.1007/JHEP12(2020)041}{{\em JHEP} {\bfseries 12}
  (2020) 041}, \href{http://arxiv.org/abs/2007.03693}{{\ttfamily
  arXiv:2007.03693 [hep-th]}}.

\bibitem{Weinberg:2003sw}
S.~Weinberg, ``{Adiabatic modes in cosmology},''
  \href{http://dx.doi.org/10.1103/PhysRevD.67.123504}{{\em Phys. Rev. D}
  {\bfseries 67} (2003) 123504},
  \href{http://arxiv.org/abs/astro-ph/0302326}{{\ttfamily
  arXiv:astro-ph/0302326}}.

\bibitem{Creminelli:2012ed}
P.~Creminelli, J.~Nore\~na, and M.~Simonovi\'c, ``{Conformal consistency
  relations for single-field inflation},''
  \href{http://dx.doi.org/10.1088/1475-7516/2012/07/052}{{\em JCAP} {\bfseries
  07} (2012) 052}, \href{http://arxiv.org/abs/1203.4595}{{\ttfamily
  arXiv:1203.4595 [hep-th]}}.

\bibitem{Hinterbichler:2012nm}
K.~Hinterbichler, L.~Hui, and J.~Khoury, ``{Conformal Symmetries of Adiabatic
  Modes in Cosmology},''
  \href{http://dx.doi.org/10.1088/1475-7516/2012/08/017}{{\em JCAP} {\bfseries
  08} (2012) 017}, \href{http://arxiv.org/abs/1203.6351}{{\ttfamily
  arXiv:1203.6351 [hep-th]}}.

\bibitem{Hinterbichler:2013dpa}
K.~Hinterbichler, L.~Hui, and J.~Khoury, ``{An Infinite Set of Ward Identities
  for Adiabatic Modes in Cosmology},''
  \href{http://dx.doi.org/10.1088/1475-7516/2014/01/039}{{\em JCAP} {\bfseries
  01} (2014) 039}, \href{http://arxiv.org/abs/1304.5527}{{\ttfamily
  arXiv:1304.5527 [hep-th]}}.

\bibitem{Pajer:2013ana}
E.~Pajer, F.~Schmidt, and M.~Zaldarriaga, ``{The Observed Squeezed Limit of
  Cosmological Three-Point Functions},''
  \href{http://dx.doi.org/10.1103/PhysRevD.88.083502}{{\em Phys. Rev. D}
  {\bfseries 88} no.~8, (2013) 083502},
  \href{http://arxiv.org/abs/1305.0824}{{\ttfamily arXiv:1305.0824
  [astro-ph.CO]}}.

\bibitem{Pajer:2020wxk}
E.~Pajer, ``{Building a Boostless Bootstrap for the Bispectrum},''
  \href{http://dx.doi.org/10.1088/1475-7516/2021/01/023}{{\em JCAP} {\bfseries
  01} (2021) 023}, \href{http://arxiv.org/abs/2010.12818}{{\ttfamily
  arXiv:2010.12818 [hep-th]}}.

\bibitem{Arkani-Hamed:2018kmz}
N.~Arkani-Hamed, D.~Baumann, H.~Lee, and G.~L. Pimentel, ``{The Cosmological
  Bootstrap: Inflationary Correlators from Symmetries and Singularities},''
  \href{http://dx.doi.org/10.1007/JHEP04(2020)105}{{\em JHEP} {\bfseries 04}
  (2020) 105}, \href{http://arxiv.org/abs/1811.00024}{{\ttfamily
  arXiv:1811.00024 [hep-th]}}.

\bibitem{Jazayeri:2021fvk}
S.~Jazayeri, E.~Pajer, and D.~Stefanyszyn, ``{From locality and unitarity to
  cosmological correlators},''
  \href{http://dx.doi.org/10.1007/JHEP10(2021)065}{{\em JHEP} {\bfseries 10}
  (2021) 065}, \href{http://arxiv.org/abs/2103.08649}{{\ttfamily
  arXiv:2103.08649 [hep-th]}}.

\bibitem{Benincasa:2022gtd}
P.~Benincasa, ``{Amplitudes meet Cosmology: A (Scalar) Primer},''
  \href{http://arxiv.org/abs/2203.15330}{{\ttfamily arXiv:2203.15330
  [hep-th]}}.

\bibitem{Senatore:2010jy}
L.~Senatore and M.~Zaldarriaga, ``{A Naturally Large Four-Point Function in
  Single Field Inflation},''
  \href{http://dx.doi.org/10.1088/1475-7516/2011/01/003}{{\em JCAP} {\bfseries
  01} (2011) 003}, \href{http://arxiv.org/abs/1004.1201}{{\ttfamily
  arXiv:1004.1201 [hep-th]}}.

\bibitem{Baumann:2015nta}
D.~Baumann, D.~Green, H.~Lee, and R.~A. Porto, ``{Signs of Analyticity in
  Single-Field Inflation},''
  \href{http://dx.doi.org/10.1103/PhysRevD.93.023523}{{\em Phys. Rev. D}
  {\bfseries 93} no.~2, (2016) 023523},
  \href{http://arxiv.org/abs/1502.07304}{{\ttfamily arXiv:1502.07304
  [hep-th]}}.

\bibitem{Baumann:2019ghk}
D.~Baumann, D.~Green, and T.~Hartman, ``{Dynamical Constraints on RG Flows and
  Cosmology},'' \href{http://dx.doi.org/10.1007/JHEP12(2019)134}{{\em JHEP}
  {\bfseries 12} (2019) 134}, \href{http://arxiv.org/abs/1906.10226}{{\ttfamily
  arXiv:1906.10226 [hep-th]}}.

\bibitem{Green:2022slj}
D.~Green, Y.~Huang, and C.-H. Shen, ``{Inflationary Adler conditions},''
  \href{http://dx.doi.org/10.1103/PhysRevD.107.043534}{{\em Phys. Rev. D}
  {\bfseries 107} no.~4, (2023) 043534},
  \href{http://arxiv.org/abs/2208.14544}{{\ttfamily arXiv:2208.14544
  [hep-th]}}.

\bibitem{Planck:2018vyg}
{\bfseries Planck} Collaboration, N.~Aghanim {\em et~al.}, ``{Planck 2018
  results. VI. Cosmological parameters},''
  \href{http://dx.doi.org/10.1051/0004-6361/201833910}{{\em Astron. Astrophys.}
  {\bfseries 641} (2020) A6}, \href{http://arxiv.org/abs/1807.06209}{{\ttfamily
  arXiv:1807.06209 [astro-ph.CO]}}. [Erratum: Astron.Astrophys. 652, C4
  (2021)].

\bibitem{Planck:2018jri}
{\bfseries Planck} Collaboration, Y.~Akrami {\em et~al.}, ``{Planck 2018
  results. X. Constraints on inflation},''
  \href{http://dx.doi.org/10.1051/0004-6361/201833887}{{\em Astron. Astrophys.}
  {\bfseries 641} (2020) A10},
  \href{http://arxiv.org/abs/1807.06211}{{\ttfamily arXiv:1807.06211
  [astro-ph.CO]}}.

\bibitem{Slosar:2019gvt}
A.~Slosar {\em et~al.}, ``{Scratches from the Past: Inflationary Archaeology
  through Features in the Power Spectrum of Primordial Fluctuations},'' {\em
  Bull. Am. Astron. Soc.} {\bfseries 51} no.~3, (2019) 98,
  \href{http://arxiv.org/abs/1903.09883}{{\ttfamily arXiv:1903.09883
  [astro-ph.CO]}}.

\bibitem{Chen:2006xjb}
X.~Chen, R.~Easther, and E.~A. Lim, ``{Large Non-Gaussianities in Single Field
  Inflation},'' \href{http://dx.doi.org/10.1088/1475-7516/2007/06/023}{{\em
  JCAP} {\bfseries 06} (2007) 023},
  \href{http://arxiv.org/abs/astro-ph/0611645}{{\ttfamily
  arXiv:astro-ph/0611645}}.

\bibitem{Chen:2008wn}
X.~Chen, R.~Easther, and E.~A. Lim, ``{Generation and Characterization of Large
  Non-Gaussianities in Single Field Inflation},''
  \href{http://dx.doi.org/10.1088/1475-7516/2008/04/010}{{\em JCAP} {\bfseries
  04} (2008) 010}, \href{http://arxiv.org/abs/0801.3295}{{\ttfamily
  arXiv:0801.3295 [astro-ph]}}.

\bibitem{Flauger:2009ab}
R.~Flauger, L.~McAllister, E.~Pajer, A.~Westphal, and G.~Xu, ``{Oscillations in
  the CMB from Axion Monodromy Inflation},''
  \href{http://dx.doi.org/10.1088/1475-7516/2010/06/009}{{\em JCAP} {\bfseries
  06} (2010) 009}, \href{http://arxiv.org/abs/0907.2916}{{\ttfamily
  arXiv:0907.2916 [hep-th]}}.

\bibitem{Flauger:2010ja}
R.~Flauger and E.~Pajer, ``{Resonant Non-Gaussianity},''
  \href{http://dx.doi.org/10.1088/1475-7516/2011/01/017}{{\em JCAP} {\bfseries
  01} (2011) 017}, \href{http://arxiv.org/abs/1002.0833}{{\ttfamily
  arXiv:1002.0833 [hep-th]}}.

\bibitem{Behbahani:2011it}
S.~R. Behbahani, A.~Dymarsky, M.~Mirbabayi, and L.~Senatore, ``{(Small)
  Resonant non-Gaussianities: Signatures of a Discrete Shift Symmetry in the
  Effective Field Theory of Inflation},''
  \href{http://dx.doi.org/10.1088/1475-7516/2012/12/036}{{\em JCAP} {\bfseries
  12} (2012) 036}, \href{http://arxiv.org/abs/1111.3373}{{\ttfamily
  arXiv:1111.3373 [hep-th]}}.

\bibitem{Behbahani:2012be}
S.~R. Behbahani and D.~Green, ``{Collective Symmetry Breaking and Resonant
  Non-Gaussianity},''
  \href{http://dx.doi.org/10.1088/1475-7516/2012/11/056}{{\em JCAP} {\bfseries
  11} (2012) 056}, \href{http://arxiv.org/abs/1207.2779}{{\ttfamily
  arXiv:1207.2779 [hep-th]}}.

\bibitem{Flauger:2014ana}
R.~Flauger, L.~McAllister, E.~Silverstein, and A.~Westphal, ``{Drifting
  Oscillations in Axion Monodromy},''
  \href{http://dx.doi.org/10.1088/1475-7516/2017/10/055}{{\em JCAP} {\bfseries
  10} (2017) 055}, \href{http://arxiv.org/abs/1412.1814}{{\ttfamily
  arXiv:1412.1814 [hep-th]}}.

\bibitem{Flauger:2016idt}
R.~Flauger, M.~Mirbabayi, L.~Senatore, and E.~Silverstein, ``{Productive
  Interactions: heavy particles and non-Gaussianity},''
  \href{http://dx.doi.org/10.1088/1475-7516/2017/10/058}{{\em JCAP} {\bfseries
  10} (2017) 058}, \href{http://arxiv.org/abs/1606.00513}{{\ttfamily
  arXiv:1606.00513 [hep-th]}}.

\bibitem{Pajer:2018egx}
E.~Pajer and D.~Stefanyszyn, ``{Symmetric Superfluids},''
  \href{http://dx.doi.org/10.1007/JHEP06(2019)008}{{\em JHEP} {\bfseries 06}
  (2019) 008}, \href{http://arxiv.org/abs/1812.05133}{{\ttfamily
  arXiv:1812.05133 [hep-th]}}.

\bibitem{Pajer:2020wnj}
E.~Pajer, D.~Stefanyszyn, and J.~Supel, ``{The Boostless Bootstrap: Amplitudes
  without Lorentz boosts},''
  \href{http://dx.doi.org/10.1007/JHEP12(2020)198}{{\em JHEP} {\bfseries 12}
  (2020) 198}, \href{http://arxiv.org/abs/2007.00027}{{\ttfamily
  arXiv:2007.00027 [hep-th]}}. [Erratum: JHEP 04, 023 (2022)].

\bibitem{Green:2022fwg}
D.~Green and Y.~Huang, ``{Flat space analog for the quantum origin of
  structure},'' \href{http://dx.doi.org/10.1103/PhysRevD.106.023531}{{\em Phys.
  Rev. D} {\bfseries 106} no.~2, (2022) 023531},
  \href{http://arxiv.org/abs/2203.10042}{{\ttfamily arXiv:2203.10042
  [hep-th]}}.

\bibitem{Hui:2022dnm}
L.~Hui, A.~Joyce, I.~Komissarov, K.~Parmentier, L.~Santoni, and S.~S.~C. Wong,
  ``{Soft theorems for boosts and other time symmetries},''
  \href{http://dx.doi.org/10.1007/JHEP02(2023)123}{{\em JHEP} {\bfseries 02}
  (2023) 123}, \href{http://arxiv.org/abs/2210.16276}{{\ttfamily
  arXiv:2210.16276 [hep-th]}}.

\bibitem{Cheung:2023qwn}
C.~Cheung, M.~Derda, A.~Helset, and J.~Parra-Martinez, ``{Soft phonon
  theorems},'' \href{http://dx.doi.org/10.1007/JHEP08(2023)103}{{\em JHEP}
  {\bfseries 08} (2023) 103}, \href{http://arxiv.org/abs/2301.11363}{{\ttfamily
  arXiv:2301.11363 [hep-th]}}.

\bibitem{Baumann:2014cja}
D.~Baumann, D.~Green, and R.~A. Porto, ``{B-modes and the Nature of
  Inflation},'' \href{http://dx.doi.org/10.1088/1475-7516/2015/01/016}{{\em
  JCAP} {\bfseries 01} (2015) 016},
  \href{http://arxiv.org/abs/1407.2621}{{\ttfamily arXiv:1407.2621 [hep-th]}}.

\bibitem{Grall:2020ibl}
T.~Grall, S.~Jazayeri, and D.~Stefanyszyn, ``{The cosmological phonon:
  symmetries and amplitudes on sub-horizon scales},''
  \href{http://dx.doi.org/10.1007/JHEP11(2020)097}{{\em JHEP} {\bfseries 11}
  (2020) 097}, \href{http://arxiv.org/abs/2005.12937}{{\ttfamily
  arXiv:2005.12937 [hep-th]}}.

\bibitem{Seery:2006vu}
D.~Seery, J.~E. Lidsey, and M.~S. Sloth, ``{The inflationary trispectrum},''
  \href{http://dx.doi.org/10.1088/1475-7516/2007/01/027}{{\em JCAP} {\bfseries
  01} (2007) 027}, \href{http://arxiv.org/abs/astro-ph/0610210}{{\ttfamily
  arXiv:astro-ph/0610210}}.

\bibitem{Pajer:2016ieg}
E.~Pajer, G.~L. Pimentel, and J.~V.~S. Van~Wijck, ``{The Conformal Limit of
  Inflation in the Era of CMB Polarimetry},''
  \href{http://dx.doi.org/10.1088/1475-7516/2017/06/009}{{\em JCAP} {\bfseries
  06} (2017) 009}, \href{http://arxiv.org/abs/1609.06993}{{\ttfamily
  arXiv:1609.06993 [hep-th]}}.

\bibitem{Bonifacio:2022vwa}
J.~Bonifacio, H.~Goodhew, A.~Joyce, E.~Pajer, and D.~Stefanyszyn, ``{The
  graviton four-point function in de Sitter space},''
  \href{http://dx.doi.org/10.1007/JHEP06(2023)212}{{\em JHEP} {\bfseries 06}
  (2023) 212}, \href{http://arxiv.org/abs/2212.07370}{{\ttfamily
  arXiv:2212.07370 [hep-th]}}.

\bibitem{Chen:2006dfn}
X.~Chen, M.-x. Huang, and G.~Shiu, ``{The Inflationary Trispectrum for Models
  with Large Non-Gaussianities},''
  \href{http://dx.doi.org/10.1103/PhysRevD.74.121301}{{\em Phys. Rev. D}
  {\bfseries 74} (2006) 121301},
  \href{http://arxiv.org/abs/hep-th/0610235}{{\ttfamily arXiv:hep-th/0610235}}.

\bibitem{Chen:2009bc}
X.~Chen, B.~Hu, M.-x. Huang, G.~Shiu, and Y.~Wang, ``{Large Primordial
  Trispectra in General Single Field Inflation},''
  \href{http://dx.doi.org/10.1088/1475-7516/2009/08/008}{{\em JCAP} {\bfseries
  08} (2009) 008}, \href{http://arxiv.org/abs/0905.3494}{{\ttfamily
  arXiv:0905.3494 [astro-ph.CO]}}.

\bibitem{Baumann:2011ws}
D.~Baumann and D.~Green, ``{A Field Range Bound for General Single-Field
  Inflation},'' \href{http://dx.doi.org/10.1088/1475-7516/2012/05/017}{{\em
  JCAP} {\bfseries 05} (2012) 017},
  \href{http://arxiv.org/abs/1111.3040}{{\ttfamily arXiv:1111.3040 [hep-th]}}.

\bibitem{Flauger:2022hie}
R.~Flauger, V.~Gorbenko, A.~Joyce, L.~McAllister, G.~Shiu, and E.~Silverstein,
  ``{Snowmass White Paper: Cosmology at the Theory Frontier},'' in {\em
  {Snowmass 2021}}.
\newblock 3, 2022.
\newblock \href{http://arxiv.org/abs/2203.07629}{{\ttfamily arXiv:2203.07629
  [hep-th]}}.

\bibitem{Weinberg:2005vy}
S.~Weinberg, ``{Quantum contributions to cosmological correlations},''
  \href{http://dx.doi.org/10.1103/PhysRevD.72.043514}{{\em Phys. Rev. D}
  {\bfseries 72} (2005) 043514},
  \href{http://arxiv.org/abs/hep-th/0506236}{{\ttfamily arXiv:hep-th/0506236}}.

\bibitem{Baumann:2022jpr}
D.~Baumann, D.~Green, A.~Joyce, E.~Pajer, G.~L. Pimentel, C.~Sleight, and
  M.~Taronna, ``{Snowmass White Paper: The Cosmological Bootstrap},'' in {\em
  {Snowmass 2021}}.
\newblock 3, 2022.
\newblock \href{http://arxiv.org/abs/2203.08121}{{\ttfamily arXiv:2203.08121
  [hep-th]}}.

\bibitem{Cabass:2016cgp}
G.~Cabass, E.~Pajer, and F.~Schmidt, ``{How Gaussian can our Universe be?},''
  \href{http://dx.doi.org/10.1088/1475-7516/2017/01/003}{{\em JCAP} {\bfseries
  01} (2017) 003}, \href{http://arxiv.org/abs/1612.00033}{{\ttfamily
  arXiv:1612.00033 [hep-th]}}.

\end{thebibliography}\endgroup

\end{document}